\documentclass[%
 reprint,
 amsmath,amssymb,
 aps,
pra,
]{revtex4-2}

\def\lL{\mathsf{L}}

\usepackage{url}
\usepackage{graphicx}
\usepackage{dcolumn}
\usepackage{bm}

\usepackage{xcolor}
\usepackage{hyperref}
\hypersetup{
    colorlinks=true,
    linkcolor=blue,
    citecolor=blue,
    filecolor=blue,      
    urlcolor=blue,
    pdftitle={Hydrodynamization at nonzero temperature},
}

\begin{document}

\title{Hydrodynamization in 1D Bose gases at nonzero temperature}

\author{Jeff Leiberton}
\author{Marcos Rigol}
\affiliation{Department of Physics, The Pennsylvania State University, University Park, Pennsylvania 16802, USA}%


\begin{abstract}
Hydrodynamization refers to the remarkably rapid process in relativistic heavy-ion collisions by which hydrodynamic descriptions become applicable. Following the observation of analogous behavior in ultracold one-dimensional (1D) Bose gases, hydrodynamization has been conjectured to be a universal dynamical phenomenon in quantum systems following high-energy quenches. Theoretical studies in this cold-atom setting have so far been restricted to quenches from ground states. Here we study how nonzero temperatures affect hydrodynamization. Specifically, using a homogeneous 1D gas of hard-core bosons, we explore how the initial temperature affects the timescales associated with hydrodynamization and prethermalization following a Bragg-pulse quench. We find that while the hydrodynamization coherence time remains unchanged, increasing temperature shortens both the damping time of the hydrodynamization oscillations and the prethermalization time. We argue that this is mainly the result of the broadening of the initial rapidity distribution, and introduce a nonzero-temperature dephasing time defined in terms of the extent of the rapidity distribution.
\end{abstract}

\maketitle


\section{Introduction}
The theory of hydrodynamics is a powerful tool for describing the long-time, large-distance evolution of generic systems far from equilibrium. Crucially, however, this theory is generally expected to apply only when the system under consideration is sufficiently close to local thermal equilibrium that its evolution can be described in terms of hydrodynamic variables and their gradients~\cite{huang_87, landau_lifshitz_59, narozhny_23}. Remarkably, in relativistic heavy-ion collisions, it was found that hydrodynamic descriptions can be applied \textit{before} the anticipated local thermalization time~\cite{bjorken_83, heller_kurkela_18, florkowski_heller_18, kurkela_taghavi_20, shen_yan_20}. This unexpectedly rapid onset of hydrodynamic behavior was called \textit{hydrodynamization} and it continues to be of much theoretical interest~\cite{strickland_24, debrito_denicol_25, lu_shi_25, abdi_nonaka_26, rajagopal_scheihing-hitschfeld_26}.

\begin{figure}[!t]
    \includegraphics[width=\linewidth]{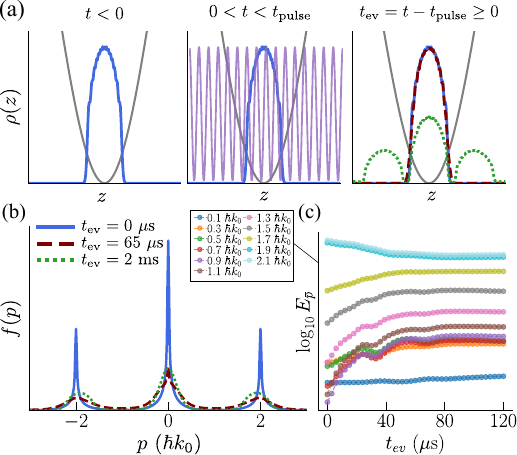}
    \vspace{-0.5cm}
    \caption{{\it Hydrodynamization in a 1D Bose gas.} (a) Illustration of the experimental sequence for a single 1D Bose gas (the experiment consists of a 2D array of 1D Bose gases)~\cite{le_zhang_23}. The gray and blue solid lines indicate the trap and the coarse-grained position distribution, respectively. At $t=0$, a Bragg pulse (purple) is applied for a time $t_\text{pulse}=10$ $\mu$s, driving the system far from equilibrium. For times $t_\text{ev}\equiv t-t_\text{pulse}>0$, the gas evolves in the original trap. (b) Momentum distributions at the evolution times shown in (a). (c) Evolution of the energy contained in different coarse-grained momentum groups. We consider the Tonks-Girardeau regime for $N=15$ $^{87}$Rb atoms in the same axial trap as in Ref.~\cite{le_zhang_23}. We take $k_0=2\pi/(775\text{ nm})$ as in the experiment.}
    \label{fig:hydrodynamization_illustration}
\end{figure}

Recently, hydrodynamization was observed experimentally in nearly integrable (quasi-)one-dimensional (1D) Bose gases \cite{le_zhang_23}. In the experiment, an array of trapped 1D Bose gases of $^{87}$Rb atoms, created using a 2D optical lattice, is driven far from equilibrium using a Bragg-pulse quench. The pulse is applied for a time $t_\text{pulse}$ and the dynamics of the gases is studied for variable times $t_\text{ev}$ after the end of the pulse. The experimental sequence is illustrated in Fig.~\ref{fig:hydrodynamization_illustration}(a) using the calculated coarse-grained position distribution of a 1D Bose gas in the Tonks-Girardeau regime. The Bragg pulse creates a coherent superposition of momentum states in which the initially occupied momenta are replicated at momenta shifted by integer multiples of $2\hbar k_0$, where $k_0=2\pi/\lambda$ and $\lambda$ is the wavelength of the Bragg standing wave; see Fig.~\ref{fig:hydrodynamization_illustration}(b). At the longest time shown in Fig.~\ref{fig:hydrodynamization_illustration}(a), a fraction of the fast-moving particles with initial momenta close to $\pm 2\hbar k_0$ have left the overlap region. Figure~\ref{fig:hydrodynamization_illustration}(b) shows that those particles slow down because of the presence of the confining potential. At short times, while all the particles are in the overlap region, this experiment can be seen as an ultracold-atom analog of a relativistic heavy-ion collision right after the colliding nuclei overlap~\cite{zhang_le_25}.

The remarkable experimental finding in Ref.~\cite{le_zhang_23} was that energy is rapidly redistributed across distant momentum modes. Specifically, as shown in Fig.~\ref{fig:hydrodynamization_illustration}(c), energy is deposited in momentum modes with $|p|\sim \hbar k_0$ on very short hydrodynamization timescales. [Note that the coarse-grained position distribution in Fig.~\ref{fig:hydrodynamization_illustration}(a) does not exhibit visible evolution during that time]. This is followed by a redistribution of particle occupations among nearby momentum modes, which results in local prethermalization. Although local prethermalization is slower than hydrodynamization, both processes occur on very short timescales that decrease with increasing $k_0$. Because of these short timescales, Ref.~\cite{le_zhang_23} conjectured that the underlying mechanism may also be relevant in nonintegrable systems, in which sufficiently long-lived quasiparticles, even if short-lived on the traditional timescales involved in low-energy dynamics, can play the role of the stable quasiparticles present at integrability. Indeed, despite the vastly different microscopic physics, a back-of-the-envelope calculation based on these ideas correctly predicted the order of magnitude of the hydrodynamization time in heavy-ion collisions~\cite{zhang_le_25}.

The theoretical analyses in Refs.~\cite{le_zhang_23, zhang_le_25}, as well as earlier theoretical studies of the Bragg-pulse quench in the Tonks-Girardeau regime~\cite{rigol_muramatsu_06, caux_bragg_16}, considered the initial pre-quench state to be a ground state. Our goal in this work is to study how nonzero temperatures affect the quantum dynamics and resulting hydrodynamization after a Bragg-pulse quench. Specifically, we explore how the initial temperature affects the damping of the hydrodynamization oscillations and the prethermalization timescale. The remainder of this paper is organized as follows. In Sec.~\ref{sec:timescales} we review the main timescales identified and discussed in Ref.~\cite{zhang_le_25}. Section~\ref{sec:model} introduces the model, the computational approach, and the observables used in this study. The results are presented in Secs.~\ref{sec:preandpostBragg} and~\ref{sec:main}. We summarize our results in Sec.~\ref{sec:summary}.

\section{Hydrodynamization and prethermalization timescales}\label{sec:timescales}

At ultracold temperatures, experimentally realized 1D Bose gases are well described by the Lieb-Liniger model with an additional confining potential term~\cite{lieb_liniger_63, yang_yang_69, cazalilla_citro_11}:
\begin{equation}
    \label{eq:LL_Ham}
    \hat{H}_{\text{LL}} = \sum_{j=1}^N \left[ -\frac{\hbar^2}{2m} \partial_{z^{}_j}^2 +U(z^{}_j) \right]+g\sum_{1\leq i<j\leq N}\delta(z^{}_i - z^{}_j),
\end{equation}
where $m$ is the mass of the atoms, $g$ is the strength of the effective 1D contact interaction, $U(z^{}_j)$ is---to a good approximation---a harmonic confining potential, and $N$ is the number of particles. The Lieb-Liniger model, corresponding to Eq.~\eqref{eq:LL_Ham} with $U(z^{}_j)=0$, is integrable. It admits stable quasiparticles whose momentum distribution, known as the rapidity distribution, determines its thermal equilibrium properties. When integrable systems are taken far from equilibrium, the conserved rapidity distribution determines the properties of the system after equilibration. The equilibrated properties are described by a generalized Gibbs ensemble (GGE)~\cite{rigol_dunjko_07, rigol_muramatsu_06, wouters_denardis_14, pozsgay_mestyan_14, ilievski_denardis_15, vidmar_rigol_16}:
\begin{equation} \label{gge_def}
\hat \rho^{}_\text{GGE} = \frac{1}{Z^{}_\text{GGE}} e^{- \! \sum_\alpha \! \lambda_\alpha \hat I_\alpha},\quad Z^{}_\text{GGE} = \text{Tr}[e^{-\! \sum_\alpha \! \lambda_\alpha \hat I_\alpha} ].
\end{equation}
Here, $\{ \hat I_\alpha \}$ is an extensive set of nontrivial conserved quantities whose existence follows from integrability. The Lagrange multipliers $\{ \lambda_\alpha \}$ are fixed so that the expectation value of each conserved quantity in the GGE matches that in the initial state,  ${\rm Tr} [ \hat \rho^{}_\text{ini} \hat I_\alpha] =  {\rm Tr} [ \hat \rho^{}_\text{GGE} \hat I_\alpha]$, where $\hat \rho^{}_\text{ini}$ is the initial density matrix.

The nearly harmonic confining potential $U(z^{}_j)$ in Eq.~\eqref{eq:LL_Ham} weakly breaks integrability. This implies that the quasiparticles are no longer stable, but they have long lifetimes compared to experimentally accessible timescales. As a result, these nearly integrable experimental systems can be locally described by the Lieb-Liniger model, and their rapidity distributions (understood as the momentum distributions of the long-lived quasiparticles) can be calculated using the local density approximation. Such theoretically calculated rapidity distributions have recently been shown to agree with the ones measured in pioneering experiments for stable ($g>0$)~\cite{wilson_malvania_20, li_zhang_23, Dubois2024Probing, dhar2025anyonization} and metastable ($g<0$)~\cite{Yang2024Phantom, horvath2025observing, bastianello_zeng_26} states in the Lieb-Liniger model. 

Near-integrability makes it possible to describe the long-time, large-distance evolution of inhomogeneous experimental systems using generalized hydrodynamics (GHD). GHD is the hydrodynamic theory of integrable systems, which is built on the assumption that, on mesoscopic scales, the system is locally described by a homogeneous GGE~\cite{castro2016emergent, bertini_collura_16, alba_bertini_21, doyon_20, essler_23, doyon_gopalakrishnan_25}. GHD has been shown to describe the dynamics of the rapidity distribution of stable~\cite{malvania_zhang_21} and metastable~\cite{Yang2024Phantom} nearly integrable 1D Bose gases following trap quenches. We stress that, because these gases are nearly integrable, the local prethermalization time (associated with the integrable theory) is well separated from the much longer thermalization time. As a result, GHD can be used to describe the dynamics over times that are long compared with the local prethermalization time but still short compared with the thermalization time.

\begin{figure}[!t]
    \includegraphics[width=\linewidth]{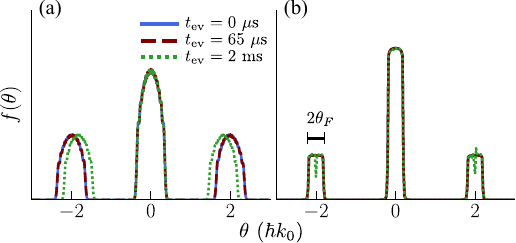}
    \vspace{-0.5cm}
    \caption{{\it Rapidity distribution in 1D Bose gases.} (a) Rapidity distributions for the same system and evolution times as in Fig.~\ref{fig:hydrodynamization_illustration}. (b) Rapidity distributions in a box trap, modeled by a homogeneous Hamiltonian with open boundary conditions, for the same evolution times as in (a). We consider the Tonks-Girardeau regime for $N=15$ $^{87}$Rb atoms, and $k_0=2\pi/(775\text{ nm})$ as in the experiment.}
    \label{fig:rapiditydistributions}
\end{figure}

In Fig.~\ref{fig:rapiditydistributions}(a), we show the rapidity distributions at the evolution times shown in Figs.~\ref{fig:hydrodynamization_illustration}(a) and~\ref{fig:hydrodynamization_illustration}(b). Note that the rapidity distribution does not change visibly during the short times over which hydrodynamization occurs. Visible changes emerge only at later times because of the confining potential. A detailed analysis in Ref.~\cite{zhang_le_25} showed that the experimental observations at short times following a Bragg-pulse quench in the presence of the confining potential are reproduced by a homogeneous system with an appropriately chosen atom density. Motivated by those results, in this work we focus on box traps. Namely, we study the dynamics under a homogeneous Hamiltonian with open boundary conditions. In Fig.~\ref{fig:rapiditydistributions}(b), we show the corresponding rapidity distributions (the momentum distributions of the underlying quasiparticles) at the evolution times shown in Fig.~\ref{fig:rapiditydistributions}(a). For this setup, the rapidity distribution is approximately conserved, with only small fluctuations arising because the open boundaries break translational invariance and mix opposite-momentum components. Hence, unlike a harmonic trap, the homogeneous nature of the box trap makes local prethermalization of the position and momentum distributions effectively equivalent (up to boundary effects) to global prethermalization.

The results shown in Figs.~\ref{fig:hydrodynamization_illustration} and~\ref{fig:rapiditydistributions} were obtained in the $g\rightarrow\infty$ limit of the Lieb-Liniger model, which is known as the Tonks-Girardeau regime. In this regime, the Hamiltonian~\eqref{eq:LL_Ham} can be mapped onto that of noninteracting spinless fermions, which play the role of the underlying quasiparticles. Consequently, all quantities shown in Figs.~\ref{fig:hydrodynamization_illustration} and~\ref{fig:rapiditydistributions} are computed exactly as explained in Sec.~\ref{sec:model}, without invoking GHD. Furthermore, for the ground state of a homogeneous noninteracting Fermi gas in the thermodynamic limit, the Fermi momentum is $\theta^{}_F = \hbar \pi \rho$, where $\rho=N/L$ is the particle density ($L$ is the length of the system). The state after the Bragg-pulse quench is known analytically in the Raman-Nath regime ($t_\text{pulse}\to0$ with fixed area under the pulse)~\cite{caux_bragg_16}. It consists of fermionic orbitals that are superpositions of plane waves with rapidities that differ by integer multiples of $2\hbar k_0$, which is consistent with the results in Fig.~\ref{fig:rapiditydistributions}.

The phases of the plane-wave components separated by $2\hbar k_0$ evolve differently in time. The characteristic energy difference $\Delta E\simeq (2\hbar k_0)^2/(2m)$ determines the frequency of the oscillations of the position-distribution modulation generated by the Bragg pulse [not shown in Fig.~\ref{fig:hydrodynamization_illustration}(a), where we plot the coarse-grained position distribution], $\omega^c_{\rm hd}=\hbar (2k_0)^2/(2m)$~\cite{le_zhang_23, zhang_le_25}. The frequency of the oscillations in Fig.~\ref{fig:hydrodynamization_illustration}(c) is $2\omega^c_{\rm hd}$ because the momentum distribution and the quantities derived from it are insensitive to the sign of the position-distribution modulation~\cite{le_zhang_23, zhang_le_25}. The period associated with $\omega^c_{\rm hd}$ is referred to as the hydrodynamization {\it coherence} time~\cite{zhang_le_25}:
\begin{equation}
    \tau_\text{hd}^{c}\equiv\frac{2\pi}{\omega_\text{hd}^{c}}=\frac{\pi m}{\hbar k_0^2}\approx 65.5\, \mu\text{s}\quad\text{for}\quad k_0=\frac{2\pi}{775\text{ nm}}.
\end{equation}
Related coherent oscillations have been studied in other contexts~\cite{deng_hagley_99, santra_baals_17, hall_yessenov_21, wu_clementi_23, bruggenjurgen_fischer_26}, where the hydrodynamization coherence time can be identified with the so-called Talbot time.

The nonvanishing width $2\theta_F$ of the occupied rapidities at zero temperature introduces a dispersion in the energy differences that leads to damping of the hydrodynamization oscillations in both position and momentum space. The bandwidth $W$ of the energy differences between the central and side peaks depends on $2\theta_F$ and on the fact that rapidity modes that are coupled differ by $\pm 2\hbar k_0$. This results in a bandwidth $W=8\hbar k_0 \theta_F/(2m)$, with a corresponding {\it dephasing} time~\cite{zhang_le_25}:
\begin{equation}
    \label{eq:tau_dp}
    \tau^{}_\text{dp}\equiv\frac{2 \pi \hbar}{W}=\frac{\pi m}{2 k_0\theta_F}.
\end{equation}
Both the damping time of the hydrodynamization oscillations and the prethermalization time depend on $\tau^{}_\text{dp}$, although their precise relation to it is strongly observable dependent, as discussed in detail in Ref.~\cite{zhang_le_25}. This will become apparent in the discussion of our results at nonzero temperature.

\section{Model and Observables}\label{sec:model}

For our calculations, we consider a lattice Tonks-Girardeau gas in the low site-occupation regime, namely, a lattice consisting of $\lL=L/a$ sites ($a$ is the lattice spacing) and $n=N/\lL\ll1$. The initial nonzero-temperature states and the dynamics after the Bragg pulse are generated by the Hamiltonian 
\begin{equation}
\label{eq:TG_Hamiltonian}
    \hat{H}=-J\sum_{j=1}^{\lL-1}(\hat{b}_{j}^{\dagger} \hat{b}^{}_{j+1}+\hat{b}_{j+1}^{\dagger}\hat{b}^{}_{j}),\quad (\hat{b}_j)^2=(\hat{b}_{j}^{\dagger})^2=0,
\end{equation}
where $J$ is the hopping integral. In the low site-occupation regime, $J$ is related to the mass and the lattice spacing via $J\simeq \hbar^2/(2ma^2)$. 

The initial density matrix is taken to have the grand-canonical form:
\begin{equation}
\label{eq:rho_GC}
    \hat{\rho}^{}_0=\frac{1}{Z^{}_0} e^{-\beta(\hat{H}-\mu\hat{N})},\quad Z^{}_0=\text{Tr}[e^{-\beta(\hat{H}-\mu\hat{N})}],
\end{equation}
where $\beta=(k^{}_BT)^{-1}$ ($k^{}_B$ is the Boltzmann constant and $T$ is the temperature), $\hat{N}=\sum_j \hat{b}_{j}^{\dagger} \hat{b}^{}_{j}$ is the total boson-number operator, and $\mu$ is the chemical potential, which is chosen so that $\text{Tr}[ \hat{N} \hat{\rho}^{}_0]=N$ with $N$ the mean number of bosons. The differences between the predictions of the canonical and grand-canonical ensembles were shown in Ref.~\cite{rigol_05} to be small even for systems with as few as $N=10$ hard-core bosons.

We then use a Bragg-pulse quench to drive the system far from equilibrium. Specifically, for times $0\leq t\leq t_\text{pulse}$, we evolve $\hat{\rho}^{}_0$ with the Hamiltonian
\begin{equation}
\label{eq:TG_Hamiltonian_Bragg}
    \hat{H}_\text{pulse}=\hat H+\sum_{j=1}^\lL U_\text{pulse}\cos^2(k_0z^{}_j) \hat{b}_{j}^{\dagger} \hat{b}^{}_{j},
\end{equation}
where $U_\text{pulse}$ is the amplitude of the applied lattice potential and $k_0=2\pi/\lambda$ is the wave number, with $\lambda=775$ nm~\cite{le_zhang_23}. For subsequent times $t_{\text{ev}}=t-t_{\text{pulse}}\geq0$, we evolve the system under the original Hamiltonian~\eqref{eq:TG_Hamiltonian} and compute the one-body density matrix
\begin{equation}
 \rho^{}_{ij}(t_\text{ev})=\text{Tr}[\hat\rho(t_\text{ev}) \hat{b}_{i}^{\dagger} \hat{b}^{}_{j}],
\end{equation}
where
\begin{align}
\hat{\rho}(t_\text{ev})&=\hat{\mathcal{U}}(t_\text{ev})\hat{\rho}^{}_{0}\,\hat{\mathcal{U}}^{\dagger}(t_\text{ev}), \quad\ \text{with}\\\hat{\mathcal{U}}(t_\text{ev})&=e^{-\frac{i}{\hbar}\hat H t_{\text{ev}}}e^{-\frac{i}{\hbar}\hat H_\text{pulse} t_{\text{pulse}}}.
\end{align}
$\hat{\rho}(t_\text{ev})$ is the time-evolving density matrix after the Bragg pulse and $\hat{\mathcal{U}}(t_\text{ev})$ is the time-evolution operator.

The initial and time-evolving one-body density matrices can be calculated exactly and efficiently by mapping the Hamiltonians~\eqref{eq:TG_Hamiltonian} and~\eqref{eq:TG_Hamiltonian_Bragg} onto those of spin-$\tfrac12$ systems and then onto those of spinless fermions using the Jordan-Wigner transformation~\cite{jordan_wigner_28, cazalilla_citro_11}. Using properties of Slater determinants, one can compute $\rho^{}_{ij}(t=0)$ for equilibrium ground states~\cite{rigol_muramatsu_04a, rigol_muramatsu_05a} and nonzero-temperature states~\cite{rigol_05}, as well as $\rho^{}_{ij}(t_\text{ev})$ far from equilibrium for initial pure~\cite{rigol_muramatsu_04b, rigol_muramatsu_05b} and nonzero-temperature~\cite{xu_rigol_17} states. The pure-state approaches from Refs.~\cite{rigol_muramatsu_04a, rigol_muramatsu_05a, rigol_muramatsu_04b, rigol_muramatsu_05b} were used to obtain the results reported in Figs.~\ref{fig:hydrodynamization_illustration} and~\ref{fig:rapiditydistributions}.

Here we study the equilibrium and far-from-equilibrium properties of the initial thermal states using a Toeplitz-like determinant approach originally developed for translationally invariant systems~\cite{lieb_schultz_04, barouch_mccoy_71}, which generalizes to general geometries. While this approach is less computationally efficient for pure states than that of Refs.~\cite{rigol_muramatsu_04a, rigol_muramatsu_05a, rigol_muramatsu_04b, rigol_muramatsu_05b}, for nonzero-temperature states it has the same favorable scaling with system size and computational stability as the approach in Refs.~\cite{rigol_05, xu_rigol_17}. Its smaller prefactor in the computational cost makes it preferable; see Appendix~\ref{sec:TD_method}. Therefore, we calculate $\rho^{}_{ij}(t)$ for initial thermal states using the expression
\begin{equation}
\label{eq:TD_representation}
\rho^{}_{ij}(t) = \frac{1}{2} \det\!\left[G^{f}_{j+\mu,\;j+\nu-1}(t) \right]_{\mu,\nu=1}^{\ell}, \ \ \ \ell=i-j>0,
\end{equation}
where the matrix entering the determinant has elements $G^f_{mn}(t) = 2 \rho_{mn}^{f}(t)-\delta_{mn}$, with $\rho_{mn}^f (t)$ denoting the equal-time one-body correlations of the corresponding spinless fermions. These correlations can be efficiently calculated as discussed in Ref.~\cite{xu_rigol_17}. For $i=j$, $\rho_{jj}^{}(t)=\rho_{jj}^f(t)$, reflecting the fact that the site occupations of hard-core bosons and spinless fermions are identical, and for $i<j$, $\rho^{}_{ij}(t)=\rho^*_{ji}(t)$. 

Let us emphasize at this point that in this work we use the terms ``ground state" or ``nonzero-temperature state'' to mean that the initial (pre-Bragg-pulse quench) state is the ground state or a nonzero-temperature state of $\hat H$~\eqref{eq:TG_Hamiltonian}, respectively. The corresponding results are labeled by $t=0$, whereas those after the Bragg-pulse quench are labeled by $t_\text{ev}=t-t_\text{pulse}\geq0$. We also stress that the states at $t=0$ ($t_\text{ev}=0$) are equilibrium (far-from-equilibrium) states of $\hat H$~\eqref{eq:TG_Hamiltonian}.

The one-body density matrix $\rho_{ij}^{}(t)$ is used to compute the momentum distribution $f(p;t)$ via a discrete Fourier transform:
\begin{equation}
\label{eq:HCB_MDF}
    f(p;t)= C\, \sum_{i,j}e^{\frac{i}{\hbar}p(z^{}_i-z^{}_j)} \rho^{}_{ij}(t),
\end{equation}
where the normalization constant $C$ is chosen so that $\int\! dpf(p;t)=1$. The momentum distribution is measured experimentally using time-of-flight imaging~\cite{le_zhang_23,bloch_dalibard_08}. 

To probe the dynamical redistribution of energy after the Bragg pulse, we compute the integrated kinetic energy per particle within a window of width $\Delta_p$ centered at $\bar{p}$, as done in Refs.~\cite{le_zhang_23, zhang_le_25}:
\begin{equation}
\label{eq:int_KE}
    E_{\bar{p}}(t_\text{ev})=\int_{\bar{p}-\Delta_p/2}^{\bar{p}+\Delta_p/2} dp f(p;t_\text{ev})\frac{p^2}{2m}.
\end{equation}
This observable highlights key dynamical features of hydrodynamization, most notably the coherent hydrodynamization oscillations, and can also diagnose regimes in which hydrodynamization is absent~\cite{zhang_le_25}.

Finally, to probe prethermalization, we compute the integrated absolute difference between the GGE prediction for the momentum distribution, which we denote by $f^{}_\text{GGE}(p)$, and the time-evolving distribution $f(p;t_\text{ev})$:
\begin{equation}
\label{eq:rel_dif_func}
    \Delta f(t_\text{ev})=\int dp \left| f(p;t_\text{ev})-f^{}_\text{GGE}(p)\right|.
\end{equation}

\section{Effect of temperature on the initial and post-Bragg-pulse states}\label{sec:preandpostBragg}

In this section, we study the effect of temperature on the initial ($t=0$, pre-Bragg-pulse) and post-Bragg-pulse ($t_\text{ev}=0$) states. For our numerical simulations, we choose the length of the system to be $L=40\times2\pi/k_0$ with open boundary conditions. We take $U_{\text{pulse}}=80E_r$, where $E_r=\hbar^2k_0^2/(2m)$ is the recoil energy, and $t_\text{pulse}=1$ $\mu$s for the Bragg-pulse quench to work in the Raman–Nath regime while keeping higher-order Bragg peaks suppressed. Unless stated otherwise, our simulations are performed on a lattice with $\lL=2000$ sites, $a=15.5$ nm, and an average of $N=15$ particles. The corresponding Fermi temperature is $T_F\approx6.45$ nK.

\begin{figure}[!t]
    \includegraphics[width=\linewidth]{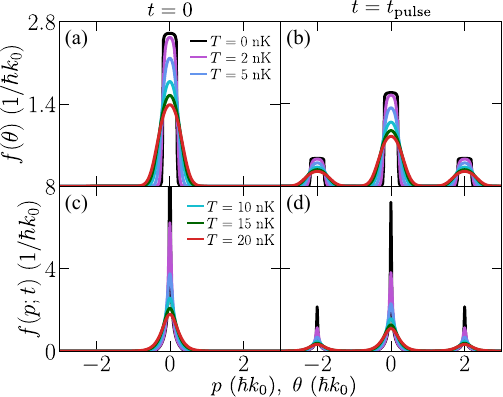}
    \vspace{-0.5cm}
    \caption{{\it Pre- and post-quench rapidity and momentum distributions.} (a),(b) Rapidity and (c),(d)
    momentum distributions in (a),(c) the initial state ($t=0$) and (b),(d) right after the Bragg-pulse quench ($t_\text{ev}=0$) for different initial temperatures. Note that the initial temperatures are chosen low enough so that all the peaks remain well resolved after the quench with negligible occupation between them.}
    \label{fig:mdf_rapidity_equil_bragg}
\end{figure}

In Fig.~\ref{fig:mdf_rapidity_equil_bragg}, we show the rapidity [Figs.~\ref{fig:mdf_rapidity_equil_bragg}(a) and~\ref{fig:mdf_rapidity_equil_bragg}(b)] and the momentum [Figs.~\ref{fig:mdf_rapidity_equil_bragg}(c) and~\ref{fig:mdf_rapidity_equil_bragg}(d)] distributions in the initial state at $t=0$  [Figs.~\ref{fig:mdf_rapidity_equil_bragg}(a) and~\ref{fig:mdf_rapidity_equil_bragg}(c)] and right after the Bragg-pulse quench at $t_\text{ev}=0$ [Figs.~\ref{fig:mdf_rapidity_equil_bragg}(b) and~\ref{fig:mdf_rapidity_equil_bragg}(d)]. As expected, increasing the temperature broadens both the rapidity [Fig.~\ref{fig:mdf_rapidity_equil_bragg}(a)] and momentum [Fig.~\ref{fig:mdf_rapidity_equil_bragg}(c)] distributions in the initial state. At nonzero temperatures, the Bragg-pulse quench produces the characteristic replicas of the initial rapidity and momentum peaks centered at integer multiples of $2\hbar k_0$ that broaden with increasing temperature. For the temperatures considered, the rapidity [Fig.~\ref{fig:mdf_rapidity_equil_bragg}(b)] and momentum [Fig.~\ref{fig:mdf_rapidity_equil_bragg}(d)] peaks remain well separated after the Bragg-pulse quench, with negligible occupation between them. Figure~\ref{fig:mdf_rapidity_equil_bragg}(d) also shows that even relatively low temperatures substantially modify the occupation of the low-momentum modes. This occurs because the power-law decay of the one-body correlations at long distances in the ground state is replaced by an exponential decay at nonzero temperatures~\cite{rigol_05}.

The broadening of the rapidity distribution with increasing temperature plays a central role in the temperature dependence of the hydrodynamization and prethermalization times as we discuss in the next section.

\section{Effect of the temperature on the quantum dynamics}\label{sec:main}

In this section, we study the effect of the initial temperature on the dynamics and timescales associated with hydrodynamization and local prethermalization. 

\begin{figure}[!t]
    \includegraphics[width=\linewidth]{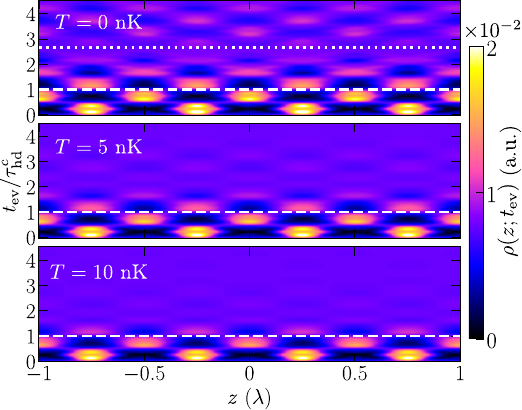}
    \vspace{-0.5cm}
    \caption{{\it Dynamics of the position distribution.} Evolution of the position distribution for three initial temperatures, $T=0$, 5, and 10 nK (from top to bottom). The Bragg pulse generates a $\lambda/2$ modulation, which oscillates with a period $\tau_\text{hd}^{c}$ (marked by the horizontal dashed lines) and whose damping increases with temperature. The horizontal dotted line in the $T=0$ panel marks the dephasing time $\tau_\text{dp}$ from Eq.~\eqref{eq:tau_dp}, defined for the initial ground state. The data shown are smoothed with a Gaussian kernel for visual clarity.}
    \label{fig:density_dynamics}
\end{figure}

As mentioned in Sec.~\ref{sec:timescales}, the Bragg-pulse quench generates a modulation of the position distribution. Because of the form of the Bragg-pulse potential [Eq.~\eqref{eq:TG_Hamiltonian_Bragg}], the wavelength of the modulation is $\lambda/2$. In Fig.~\ref{fig:density_dynamics}, we plot the evolution of the position distribution $\rho(z;t_\text{ev})$ at the center of the system (within four wavelengths of the density modulation) for three initial temperatures. At early times, the position distribution exhibits clear oscillations with period $\tau_\text{hd}^{c}$ (marked by the horizontal dashed lines) at all three temperatures. In the top panel ($T=0$), we also mark the dephasing time $\tau_\text{dp}$ [Eq.~\eqref{eq:tau_dp}] with a horizontal dotted line. The density modulation becomes noticeably blurred near the dephasing time, and this is followed by a partial revival of the oscillations, which gradually damp out over the time interval shown. Increasing the temperature enhances the damping, so that at the highest temperature shown ($T=10$ nK) the oscillations are strongly suppressed beyond the first cycle.

The Tonks-Girardeau limit of the Lieb-Liniger model is special because observables that are {\it diagonal} in position space, such as the position distribution in Fig.~\ref{fig:density_dynamics}, are identical to those of a system of noninteracting spinless fermions. In contrast, observables that are {\it off-diagonal} in position space, such as the momentum distribution, differ markedly between hard-core bosons and noninteracting fermions~\cite{rigol_muramatsu_04a, rigol_muramatsu_05a}. For example, compare Figs.~\ref{fig:hydrodynamization_illustration}(b) and~\ref{fig:rapiditydistributions}(a), and the top and bottom panels of Fig.~\ref{fig:mdf_rapidity_equil_bragg}. Following the Bragg pulse, the rapidity distribution (the momentum distribution of the noninteracting fermions) does not evolve significantly in our setup. It exhibits only small oscillations because of the open boundary conditions; see Fig.~\ref{fig:rapiditydistributions}(b). For periodic boundary conditions, the rapidity distribution is conserved. By contrast, the momentum distribution $f(p;t)$ of the Tonks-Girardeau gas does evolve significantly because of the interactions. The evolution of $f(p;t)$ depends strongly on momentum, and the timescales associated with hydrodynamization and local prethermalization are reflected differently in $f(p;t)$ and in the integrated kinetic energy per particle $E_{\bar{p}}(t_\text{ev})$ defined in Eq.~\eqref{eq:int_KE}.

\begin{figure}[!t]
    \includegraphics[width=\linewidth]{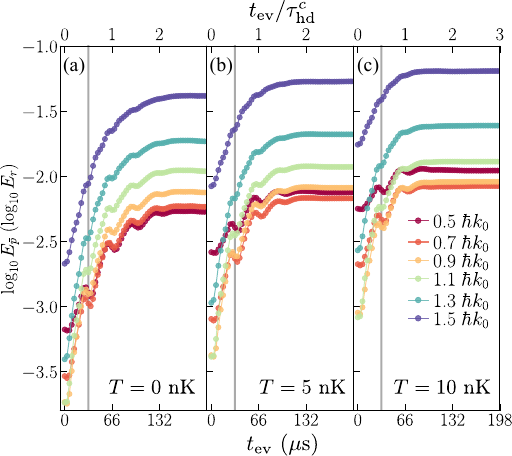}
    \vspace{-0.5cm}
    \caption{\textit{Hydrodynamization oscillations.} Evolution of the integrated kinetic energy per particle $E_{\bar{p}}(t_\text{ev})$ [see Eq.~\eqref{eq:int_KE}] for three initial temperatures: (a) $T=0$ nK, (b) $T=5$ nK, and (c) $T=10$ nK. We show results for $\bar{p}/(\hbar k_0)=0.5$--1.5, and the momentum-window width is $\Delta_p=0.2\hbar k_0$. The upper axis shows the evolution time in units of the hydrodynamization coherence time $\tau_\text{hd}^{c}$. The vertical gray line marks $t_\text{ev}=\tau_\text{hd}^{c}/2\approx 33$~$\mu$s, corresponding to the characteristic period of the hydrodynamization oscillations in $E_{\bar{p}}(t_\text{ev})$.}
    \label{fig:int_KE}
\end{figure}

The effect of temperature on hydrodynamization is reflected in how energy is deposited into momentum modes around $\hbar k_0$. In Fig.~\ref{fig:int_KE}, we show the short-time evolution of $E_{\bar{p}}(t_\text{ev})$, with $\Delta_p/(\hbar k_0)=0.2$, focusing on $\bar{p}$ between $0.5\hbar k_0$ and $1.5\hbar k_0$. We show results for three initial temperatures: 0 nK [Fig.~\ref{fig:int_KE}(a)], 5 nK [Fig.~\ref{fig:int_KE}(b)], and 10 nK [Fig.~\ref{fig:int_KE}(c)]. The energies associated with these momentum modes increase rapidly at all temperatures and exhibit hydrodynamization oscillations with a momentum- and temperature-independent period $\tau_\text{hd}^{c}/2$. The initial temperature has two main effects. As expected, higher temperatures increase the energy present in intermediate momentum modes at $t_\text{ev}=0$. More importantly, higher temperatures increase the damping of the hydrodynamization oscillations, resulting in fewer visible oscillations in $E_{\bar{p}}(t_\text{ev})$.

\begin{figure}[!t]
    \includegraphics[width=\linewidth]{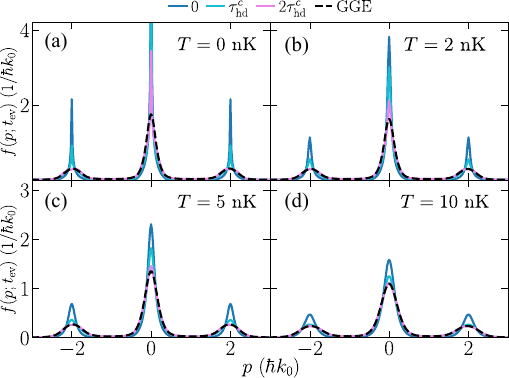}
    \vspace{-0.5cm}
    \caption{{\it Dynamics of the momentum distribution.} Evolution of the momentum distribution for four initial temperatures: (a) $T=0$ nK, (b) $T=2$ nK, (c) $T=5$ nK, and (d) $T=10$ nK. We show the momentum distributions at times $t_\text{ev}/\tau_\text{hd}^{c}= 0,\,1,\,2$, together with the corresponding GGE predictions.}
    \label{fig:late_time_MDF}
\end{figure}

Prethermalization and its temperature dependence are reflected in the dynamics of the entire momentum distribution. In Fig.~\ref{fig:late_time_MDF}, we show how the momentum distributions evolve for four initial temperatures: 0 nK [Fig.~\ref{fig:late_time_MDF}(a)], 2 nK [Fig.~\ref{fig:late_time_MDF}(b)], 5 nK [Fig.~\ref{fig:late_time_MDF}(c)], and 10 nK [Fig.~\ref{fig:late_time_MDF}(d)]. Specifically, we show the momentum distributions at $t_\text{ev}/\tau_\text{hd}^{c}= 0,\,1,\,2$ and compare them with the corresponding GGE predictions for each state generated by the Bragg-pulse quench. The GGE is constructed using the occupations of the eigenstates of the corresponding noninteracting fermions in the box trap. In contrast to the rapidity distributions shown in Fig.~\ref{fig:rapiditydistributions}, these occupations are exactly conserved. The results in Fig.~\ref{fig:late_time_MDF} show that, independently of the initial temperature, the momentum distribution $f(p;t)$ away from integer multiples of $2\hbar k_0$ becomes indistinguishable from the GGE predictions more rapidly than the momentum distribution $f(p;t)$ close to integer multiples of $2\hbar k_0$~\cite{le_zhang_23, zhang_le_25}. The results in Fig.~\ref{fig:late_time_MDF} also show that with increasing temperature prethermalization of the entire momentum distribution occurs in shorter times.

\begin{figure}[!t]
    \includegraphics[width=\linewidth]{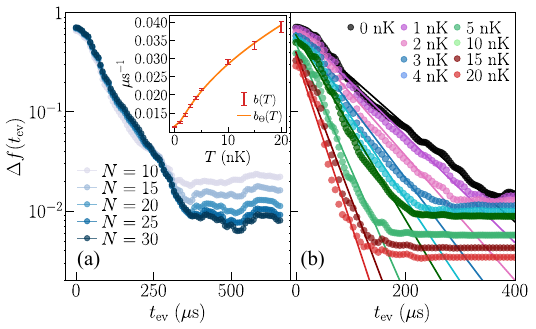}
    \vspace{-0.5cm}
    \caption{{\it Local prethermalization.} (a) $\Delta f(t_\text{ev})$ for initial ground states with $N=10,\, 15,\,20,\,25,\,30$ at fixed average density $N/L$. (b) $\Delta f(t_\text{ev})$ for initial states with different temperatures and $N=15$. The straight lines show fits of early-time results to $a(T)e^{-b(T)t_\text{ev}}$. Inset in (a): Results for $b(T)$ obtained from the fits in panel (b) and $b^{}_\Theta(T)$ obtained by evaluating Eq.~\eqref{eq:bnonzeroT} using $\Theta(T)$ from Eq.~\eqref{eq:ThetanonzeroT}.}
    \label{fig:deltaf}
\end{figure}

Next, we study $\Delta f(t_\text{ev})$ [see Eq.~\eqref{eq:rel_dif_func}] to quantify the effect of temperature on the prethermalization dynamics. Before analyzing the temperature dependence, we verify that finite-size effects (expected to be strongest for initial ground states) do not appreciably affect the extraction of the early-time decay rate. In Fig.~\ref{fig:deltaf}(a) we plot $\Delta f(t_\text{ev})$ for initial ground states with different $N$ and $L$ but fixed average density $N/L=0.075/a$. For small systems, finite-size effects cause $\Delta f(t_\text{ev})$ to develop a late-time plateau with a nonzero value that decreases with system size. For the largest systems shown in Fig.~\ref{fig:deltaf}(a), one can further see that the plateau is replaced by the onset of a power-law decay of $\Delta f(t_\text{ev})$ that was studied in detail in Ref.~\cite{zhang_le_25}. Hence, in agreement with the conclusions of Ref.~\cite{zhang_le_25}, we expect $\Delta f(t_\text{ev})\rightarrow 0$ at long times in the thermodynamic limit. More importantly for our study, for the system sizes shown in the plot, the early-time near-exponential decay of $\Delta f(t_\text{ev})$ is not appreciably affected by finite-size effects. Therefore, using finite-size results for $\Delta f(t_\text{ev})$ we can determine the decay rate. 

In Fig.~\ref{fig:deltaf}(b) we show our results for $\Delta f(t_\text{ev})$ at different initial temperatures. The results for $T>0$ are qualitatively similar to those for $T=0$, with the main difference being that $\Delta f(t_\text{ev},T)\propto e^{-b(T) t_\text{ev}}$ decays (at short times) at a faster rate $b(T)$ with increasing $T$. This is quantified by the fits to early-time data shown in Fig.~\ref{fig:deltaf}(b). The values $b(T)$ obtained using those fits are shown in the inset in Fig.~\ref{fig:deltaf}(a). The main effect of temperature on $b(T)$ can be understood as resulting from the broadening of the rapidity distribution. In Ref.~\cite{zhang_le_25} it was shown that, at zero temperature
\begin{equation}\label{eq:bzeroT}
 b(T=0)=\frac{C}{\tau^{}_\text{dp}}=\frac{C\, 2 k_0 \theta_F}{m\pi}= C' \theta_F,   
\end{equation}
where $C$ is a constant that in general depends on the system and Bragg-pulse parameters (for our system we determine it by fitting the $T=0$ results), and $C'\equiv C\, 2 k_0/(m\pi)$. We can generalize Eq.~\eqref{eq:bzeroT} to nonzero temperature as
\begin{equation}\label{eq:bnonzeroT}
 b^{}_\Theta(T)= C' \Theta(T),   
\end{equation}
where $\Theta(T)$ characterizes the extent of the rapidity distribution at temperature $T$, reducing to $\theta_F$ in the limit $T\rightarrow0$. For the homogeneous system considered here, a natural family of choices satisfying this constraint is 
\begin{equation}\label{eq:DeltanonzeroT}
\Delta_{2w}(T)=\left[(2w+1) \int_{-\infty}^\infty d\theta\, \theta^{2w} f(\theta;T)\right]^{\frac{1}{2w}}.
\end{equation}
The factor $(2w+1)^{\frac{1}{2w}}$ is chosen so that $\Delta_{2w}(T)$ reduces to the Fermi momentum $\theta_F$ for the zero-temperature rapidity distribution in the thermodynamic limit. We find that $2w=12$ provides the best description of our fitted values of $b(T)$, so we take 
\begin{equation}\label{eq:ThetanonzeroT}
\Theta(T)\equiv \Delta_{12}(T)=\left[13 \int_{-\infty}^\infty d\theta\, \theta^{12} f(\theta;T)\right]^{\frac{1}{12}}.
\end{equation}
In the inset in Fig.~\ref{fig:deltaf}(a) we show our results for $b^{}_\Theta(T)$ in Eq.~\eqref{eq:bnonzeroT} with $\Theta(T)$ in Eq.~\eqref{eq:ThetanonzeroT} evaluated numerically in our finite chains after replacing the open boundary conditions by periodic ones. This is done to eliminate the spurious effect of the high-rapidity tails generated by the open boundary conditions in our finite chains. The calculated value of $b^{}_\Theta(T)$ provides a good description of $b(T)$ at all the temperatures shown.

The observed agreement between $b^{}_\Theta(T)$ and $b(T)$ suggests that we can define a nonzero-temperature dephasing time that generalizes Eq.~\eqref{eq:tau_dp} as:
\begin{equation}
    \label{eq:tau_dp_T}
    \tau^{}_\text{dp}(T)\equiv\frac{\pi m}{2 k_0 \Theta(T)}.
\end{equation}
This nonzero-temperature dephasing time is expected to control both the damping time of the hydrodynamization oscillations and the prethermalization time, with the precise relation between the latter two and $\tau^{}_\text{dp}(T)$ being strongly observable-dependent.

\begin{figure}[!t]
    \includegraphics[width=\linewidth]{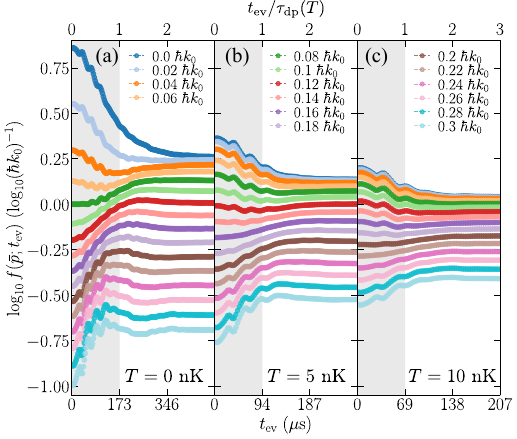}
    \vspace{-0.5cm}
    \caption{\textit{Hydrodynamization and local prethermalization.} Evolution of $f(\bar p;t)$, with $\bar{p}\in [0,0.3]\hbar k_0$, for three initial temperatures: (a) $T=0$ nK, (b) $T=5$ nK, and (c) $T=10$ nK. The upper axis shows the evolution time in units of the nonzero-temperature dephasing time $\tau^{}_\text{dp}(T)$; see Eq.~\eqref{eq:tau_dp_T}. The gray shaded region marks times $t_\text{ev}<\tau^{}_\text{dp}(T)$.}
    \label{fig:fpbar}
\end{figure}

In Fig.~\ref{fig:fpbar}, we show the evolution of $f(\bar p;t)$, with $\bar{p}\in [0,0.3]\hbar k_0$, for three initial temperatures: 0 nK [Fig.~\ref{fig:fpbar}(a)], 5 nK [Fig.~\ref{fig:fpbar}(b)], and 10 nK [Fig.~\ref{fig:fpbar}(c)]. Note that the time interval shown shortens with increasing $T$. However, it is the same in all panels when expressed in units of $\tau^{}_\text{dp}(T)$ (see the upper axis). Hydrodynamization oscillations are apparent at all temperatures, and are largely damped within one dephasing time $\tau^{}_\text{dp}(T)$. Local prethermalization, signaled by the relaxation of the results to nearly constant values, is also apparent at all temperatures. Note that we show results for low momenta $\bar p \in [0,0.3]\hbar k_0$, for which local prethermalization is slowest. While our calculations for $\Delta f(t_\text{ev})$ provide a single characteristic prethermalization time for the momentum distribution, Fig.~\ref{fig:fpbar}(a) shows that the local prethermalization of $f(p;t)$ involves a range of timescales that become longer as $\bar{p}$ decreases. This was discussed in detail in Ref.~\cite{zhang_le_25}. Comparing Figs.~\ref{fig:fpbar}(a)--\ref{fig:fpbar}(c), we see that the second major effect of increasing temperature, in addition to the decrease of $\tau^{}_\text{dp}(T)$ already discussed, is that it reduces the range of local prethermalization times, i.e., the difference between the prethermalization times of the high- and low-momentum modes. As the temperature increases, both the damping of the hydrodynamization oscillations and the prethermalization of momentum modes across the entire momentum distribution tend to occur within one dephasing time $\tau^{}_\text{dp}(T)$. 

\section{Summary and discussion}\label{sec:summary}

In this work, we investigated the role of temperature on the dynamics of 1D Bose gases following Bragg-pulse quenches. We found that the phenomenology of hydrodynamization and local prethermalization discussed in the context of quenches from initial ground states in Refs.~\cite{le_zhang_23, zhang_le_25} is robust against nonzero temperature, providing further evidence that they are universal dynamical phenomena following high-energy quenches. While temperature does not affect the hydrodynamization coherence time $\tau_\text{hd}^c$, it does affect the damping of the hydrodynamization oscillations and local prethermalization. Its primary effect is to broaden the rapidity distribution, thereby shortening the temperature-dependent dephasing time $\tau_\text{dp}(T)$, which in turn controls the damping of the hydrodynamization oscillations and local prethermalization. We argued that $\tau_\text{dp}(T)$ can be defined in terms of the extent of the initial rapidity distribution.

After accounting for the shortening of the dephasing time $\tau_\text{dp}(T)$, we identified a second effect of temperature, clearly seen in the progression of Figs.~\ref{fig:fpbar}(a)--\ref{fig:fpbar}(c). Local prethermalization of the momentum distribution for initial ground states consists of a range of local prethermalization times, which become longer as the momentum decreases~\cite{le_zhang_23, zhang_le_25}. With increasing temperature we found that the range of local prethermalization times decreases, resulting in the prethermalization of all momentum modes occurring within one dephasing time $\tau^{}_\text{dp}(T)$. 

The latter findings can be understood as a consequence of the faster decay of one-body correlations with increasing temperature, which primarily affects the occupation of low-momentum modes in the initial state. This suggests a broader physical picture: initial states with short-range correlations--whether ground states or nonzero-temperature states--may exhibit local prethermalization across the entire momentum distribution at approximately the same time. For the systems studied here, local prethermalization occurs together with the damping of the hydrodynamization oscillations within one dephasing time $\tau^{}_\text{dp}(T)$. The observed quantitative differences between the experimental results at finite $\gamma$ and their theoretical modeling in the $\gamma=\infty$ Tonks-Girardeau regime~\cite{le_zhang_23, zhang_le_25} may be related to the longer-ranged one-body correlations in the former. In the experiments, the prethermalization time was found to be about three times longer at a low (so-called $p_{50}$) momentum. Since decreasing $\gamma$ leads to a slower decay of the ground-state one-body correlations, it may therefore also lead to longer prethermalization times. Exploring this question will be the subject of future work.

\begin{acknowledgments}
We thank David S. Weiss and Yicheng Zhang for many stimulating discussions about hydrodynamization and local prethermalization in 1D Bose gases. We acknowledge the support of the National Science Foundation (NSF) Grant No.~PHY-2309146. This work used Anvil at Purdue University through allocation PHY-260071 from the Advanced Cyberinfrastructure Coordination Ecosystem: Services \& Support (ACCESS) program, which is supported by U.S. NSF grants \#2138259, \#2138286, \#2138307, \#2137603, and \#2138296.
\end{acknowledgments}

\appendix

\section{Determinantal Calculation of Equal-Time Correlations}
\label{sec:TD_method}

In this Appendix, we outline a general derivation of Eq.~\eqref{eq:TD_representation}, which holds for particle-number-conserving fermionic Gaussian states described by a density matrix $\hat{\rho}^f_G$. Such states are fully characterized by their one-body density matrix
\begin{equation}
    \rho_{mn}^f=\text{Tr}\left[\hat\rho^f_G \hat{c}_{m}^{\dagger} \hat{c}^{}_{n}\right],
\end{equation}
where $\hat{c}_{m}^{\dagger}$ and $\hat{c}_{m}^{}$ are, respectively, the creation and annihilation operators for a spinless fermion at site $m$. In the main text, three types of particle-number-conserving fermionic Gaussian states underlie our hard-core boson calculations: nonzero-temperature initial states $\hat{\rho}^{f}_{0}$, time-evolving states $\hat{\rho}^f(t_\text{ev})$, and GGEs $\hat \rho^{f}_\text{GGE}$.

In what follows, we use the notation
\begin{equation}
    \langle \hat {\mathcal{O}} \rangle^{}_{G}\equiv \text{Tr}\left[\hat{\rho}^{f}_{G} \hat{\mathcal{O}} \right]
\end{equation}
to denote the expectation value of an operator $\hat{\mathcal{O}}$ in the fermionic Gaussian state $\hat{\rho}^{f}_{G}$. 

Our goal is to prove that, for $\ell>0$, the hard-core-boson one-body density matrix satisfies
\begin{align}
    \rho_{j+\ell,j}^{}&\equiv\langle\hat b_{j+\ell}^\dagger\hat b^{}_{j}\rangle^{}_G =\left\langle \hat{c}_{j+\ell}^{\dagger}e^{-i\pi\hat{\theta}^{}_{j+\ell,j}}\hat{c}^{}_{j}\right\rangle^{}_{G} \label{eq:prove1} \\&=\frac{1}{2}\det\left[G_{j+\mu,\;j+\nu-1}^{f}\right]_{\mu,\nu=1}^{\ell},\label{eq:prove2}
\end{align}
where $\hat{\theta}_{m,n}=\sum_{m> \gamma > n}\hat{n}_{\gamma}^{}$ (for $m>n$), $\hat{n}_{\gamma}^{}=\hat{c}_{\gamma}^\dagger \hat{c}^{}_{\gamma}$, and $G_{mn}^{f}=2 \rho_{mn}^{f}-\delta_{mn}$. 

Equation~\eqref{eq:prove1} follows straightforwardly from the Jordan-Wigner transformation
\begin{equation}
    \hat{b}_{j}^{\dagger}=\hat{c}_{j}^{\dagger}e^{-i\pi\hat{\theta}^{}_{j,0}}, \qquad \hat{b}_{j}=e^{i\pi\hat{\theta}^{}_{j,0}}\hat{c}^{}_{j},
\end{equation}
and the identity $e^{-i\pi\hat{n}_{j}}\hat{c}_{j} = \hat{c}_{j}$. We then expand 
\begin{align}
     e^{-i\pi\hat{\theta}_{j+\ell,j}}&=\prod_{0<\gamma<\ell}\left(1-2\,\hat{n}_{j+\gamma}^{} \right) \label{eq:prod_exp1} \\ &=\sum_{X\in\mathcal{P}(\Gamma)}(-2)^{|X|}\prod_{x\in X} \hat{n}_{j+x}^{} \label{eq:prod_exp2},
\end{align}
where $\Gamma=\{ 1,2,\ldots,\ell-1 \}$ has cardinality $|\Gamma|=\ell-1$, and $\mathcal{P}(\Gamma)$ denotes its power set. We use the conventions $|\emptyset|=0$ and $\prod_{x\in\emptyset} \hat{n}_{j+x}^{}=\hat{I}$. For nonempty subsets $X$, the product is taken in increasing order of $x$ (its value is independent of the ordering because the number operators commute). We obtain Eq.~\eqref{eq:prod_exp2} by expanding the product in Eq.~\eqref{eq:prod_exp1} and associating each term with a subset $X\in\mathcal{P}(\Gamma)$.

Inserting Eq.~\eqref{eq:prod_exp2} into Eq.~\eqref{eq:prove1}, we find
\begin{align}
    \label{eq:LHS_expand} \rho^{}_{j+\ell,j}&=\sum_{X\in\mathcal{P}\left(\Gamma\right)}\left(-2\right)^{\left|X\right|}\left\langle \hat{c}_{j+\ell}^{\dagger}\left(\prod_{x\in X}\hat{n}^{}_{j+x}\right)\hat{c}_{j}^{}\right\rangle ^{}_{G}\\&=\sum_{X\in\mathcal{P}\left(\Gamma\right)}2^{\left|X\right|}\det\left[\rho_{j+\mu,j+\nu-1}^{f}\right]_{\substack{\mu\in\left(X,\ell\right)\\ \nu-1\in\left(0,X\right)}},\label{eq:RHS_expand}
\end{align}
where $(X,\ell)$ denotes the ordered set of all $x\in X$ and $\ell$, and $(0,X)$ denotes the ordered set of 0 and all $x\in X$. To obtain the second line, we used the fact that, for $X=\{x^{}_1<x^{}_2<\cdots<x^{}_{|X|}\}$, Wick's theorem gives
\begin{equation}
\left\langle\!\! \hat c_{j+\ell}^{\dagger} \! \left(\prod_{x\in X}\hat n_{j+x}\!\right)\! \hat c^{}_j \!
\right\rangle^{}_{\!\!G}\!\! = (-1)^{|X|} \det \!\left[ \rho_{j+\mu,j+\nu-1}^{f} \right]_{\!\!\!\substack{\mu\in(X,\ell)\\
\nu-1\in(0,X)}}\!\!\!.
\end{equation}
Note that the factor of $(-1)^{|X|}$ appearing in Eq.~\eqref{eq:LHS_expand} is canceled by the sign $(-1)^{|X|}$ arising from the fermionic reordering needed to apply Wick's theorem. Wick's theorem expresses each $2(|X|+1)$-point correlation function as a signed sum of products of $|X|+1$ two-point correlation functions. In the present $U(1)$-symmetric case, this sum can be written as a determinant~\cite{surace_tagiliacozzo_22}.

Now we can show that Eq.~\eqref{eq:prove2} reproduces Eq.~\eqref{eq:RHS_expand}. For a given $j$ and $\ell$, the matrix whose determinant appears in Eq.~\eqref{eq:prove2} can be written as 
\begin{widetext}
\begin{equation}
   \left[G_{j+\mu,\;j+\nu-1}^{f}\right]_{\mu,\nu=1}^{\ell} = \left(\begin{array}{c|c|c|c}
2\boldsymbol{\rho}_{j}^{f} & 2\boldsymbol{\rho}_{j+1}^{f}-\boldsymbol{\delta}^{}_{1} & \cdots & 2\boldsymbol{\rho}_{j+\ell-1}^{f}-\boldsymbol{\delta}^{}_{\ell-1}\end{array}\right)_{\ell\times\ell},
\end{equation}
where $\left(\boldsymbol{\rho}^{f}_{j+n}\right)_{m}=\rho_{j+m,j+n}^{f}$ and $(\boldsymbol{\delta}^{}_{n})_m=\delta_{mn}$ are column vectors of size $\ell$ with $1\leq m \leq \ell$ and $0\leq n \leq \ell-1$. By the multilinearity of the determinant, we can successively expand each of the last $\ell -1$ columns. For example, expanding the second column, we find
\begin{equation}
\label{eq:multilin_expand}
\det\left[G_{j+\mu,\;j+\nu-1}^{f}\right]_{\mu,\nu=1}^{\ell}=\det\left(\begin{array}{c|c|c|c}
2\boldsymbol{\rho}_{j}^{f} & 2\boldsymbol{\rho}_{j+1}^{f} & \cdots & 2\boldsymbol{\rho}_{j+\ell-1}^{f}-\boldsymbol{\delta}^{}_{\ell-1}\end{array}\right)+\det\left(\begin{array}{c|c|c|c}
2\boldsymbol{\rho}_{j}^{f} & -\boldsymbol{\delta}^{}_{1} & \cdots & 2\boldsymbol{\rho}_{j+\ell-1}^{f}-\boldsymbol{\delta}^{}_{\ell-1}\end{array}\right).
\end{equation}
Repeating this procedure for the remaining $\ell-2$ columns produces a sum of $2^{\ell-1}$ determinants. The resulting expression can be simplified by expanding each determinant along its $-\boldsymbol{\delta}_n$ columns using the cofactor formula. Specifically, each term in the multilinear expansion contains $V$ columns $\{ -\boldsymbol{\delta}^{}_{n_{\alpha}} \}_{\alpha=1}^{V}$, where $0 \leq V \leq \ell -1$. Expanding the determinant successively along these columns, in increasing order of their original column labels, results in a single determinant of size $\ell-V$ with a positive sign. To see this, consider the next $-\boldsymbol{\delta}$ column to be expanded after any number of successive eliminations. If its nonzero entry $-1$ lies in row $m'$ of the reduced matrix, then the column is in position $m'+1$. The corresponding cofactor contributes a sign $(-1)^{m'+(m'+1)}=-1$, which cancels the minus sign of that entry. Proceeding in this way for each term, we can write
\begin{align}
    \det\left[G_{j+\mu,j+\nu-1}^{f}\right]_{\mu,\nu=1}^{\ell}&=2^{\ell}\det\left[\rho_{j+\mu,j+\nu-1}^{f}\right]_{\mu,\nu=1}^{\ell}+2^{\ell-1}\sum_{1\leq m\leq\ell-1}\det\left[\rho_{j+\mu,j+\nu-1}^{f}\right]_{\mu\neq m,\nu-1\neq m}^{\ell-1}\\&\quad +2^{\ell-2}\sum_{1\leq m<n\leq\ell-1}\det\left[\rho_{j+\mu,j+\nu-1}^{f}\right]_{\mu\notin\left\{m,n\right\},\nu-1\notin\left\{m,n\right\}}^{\ell-2}+\ldots+2\rho_{j+\ell,j}^{f}.
\end{align}

Finally, each term in the above expression can be associated with a subset $X\in \mathcal{P}(\Gamma)$. For a given subset $X$, the columns among the second through the $\ell$th columns that contain vectors $2\boldsymbol{\rho}_{j+x}^{f}$ are precisely those labeled by $x\in X$. The remaining columns, labeled by $x'\notin X$, contain $-\boldsymbol{\delta}_{x'}$. For example, the following associations hold:
\begin{align}
    2^\ell \det\left[\rho_{j+\mu,j+\nu-1}^{f}\right]_{\mu,\nu=1}^{\ell} & \longleftrightarrow \Gamma\\
    2^{\ell-1}\det\left[\rho_{j+\mu,j+\nu-1}^{f}\right]_{\mu\neq 1,\nu-1\neq 1}^{\ell-1} & \longleftrightarrow \Gamma\setminus\{1\} \\ 
    \vdots \nonumber \\
    2 \rho_{j+\ell,j}^{f} & \longleftrightarrow \emptyset. 
\end{align}
These associations yield
\begin{equation}
    \frac{1}{2}\det\left[G_{j+\mu,\;j+\nu-1}^{f}\right]_{\mu,\nu=1}^{\ell}=\sum_{X\in\mathcal{P}\left(\Gamma\right)}2^{\left|X\right|}\det\left[\rho_{j+\mu,j+\nu-1}^{f}\right]_{\substack{\mu\in\left(X,\ell\right)\\
\nu-1\in\left(0,X\right)}},
\end{equation}
\end{widetext}
which establishes Eq.~\eqref{eq:prove2}.

We end this Appendix with a few remarks. The direct calculation of all $O(\lL^2)$ matrix elements of the one-body density matrix of hard-core bosons involves determinants of matrices whose linear dimensions are at most $O(\lL)$. Since evaluating the determinant of a matrix of linear dimension $O(\lL)$ requires $O(\lL^3)$ operations, the overall computational cost scales as $O(\lL^5)$. This is the same scaling as the approaches in Refs.~\cite{rigol_05, xu_rigol_17}. However, the present approach has a smaller prefactor because the matrices whose determinants must be evaluated are smaller or equal to those in Refs.~\cite{rigol_05, xu_rigol_17}.

For translationally invariant systems, the matrix $G_{mn}^{f}$ depends only on the difference $m-n$. Consequently, $[G_{j+\mu,j+\nu-1}^{f}]_{\mu,\nu=1}^{\ell}$ has a Toeplitz structure: its $(\mu,\nu)$ entry depends only on $\mu-\nu$. Toeplitz determinants have been studied extensively in the mathematical physics literature~\cite{grenander_szego_58, mccoy_wu_68, bottcher_silbermann_99}. The asymptotic properties of large Toeplitz determinants have attracted
particular attention, as they can be used, for example, to derive the decay
of correlation functions analytically~\cite{barouch_mccoy_71}.

\bibliography{main}

\begin{thebibliography}{58}%
\makeatletter
\providecommand \@ifxundefined [1]{%
 \@ifx{#1\undefined}
}%
\providecommand \@ifnum [1]{%
 \ifnum #1\expandafter \@firstoftwo
 \else \expandafter \@secondoftwo
 \fi
}%
\providecommand \@ifx [1]{%
 \ifx #1\expandafter \@firstoftwo
 \else \expandafter \@secondoftwo
 \fi
}%
\providecommand \natexlab [1]{#1}%
\providecommand \enquote  [1]{``#1''}%
\providecommand \bibnamefont  [1]{#1}%
\providecommand \bibfnamefont [1]{#1}%
\providecommand \citenamefont [1]{#1}%
\providecommand \href@noop [0]{\@secondoftwo}%
\providecommand \href [0]{\begingroup \@sanitize@url \@href}%
\providecommand \@href[1]{\@@startlink{#1}\@@href}%
\providecommand \@@href[1]{\endgroup#1\@@endlink}%
\providecommand \@sanitize@url [0]{\catcode `\\12\catcode `\$12\catcode `\&12\catcode `\#12\catcode `\^12\catcode `\_12\catcode `\%12\relax}%
\providecommand \@@startlink[1]{}%
\providecommand \@@endlink[0]{}%
\providecommand \url  [0]{\begingroup\@sanitize@url \@url }%
\providecommand \@url [1]{\endgroup\@href {#1}{\urlprefix }}%
\providecommand \urlprefix  [0]{URL }%
\providecommand \Eprint [0]{\href }%
\providecommand \doibase [0]{https://doi.org/}%
\providecommand \selectlanguage [0]{\@gobble}%
\providecommand \bibinfo  [0]{\@secondoftwo}%
\providecommand \bibfield  [0]{\@secondoftwo}%
\providecommand \translation [1]{[#1]}%
\providecommand \BibitemOpen [0]{}%
\providecommand \bibitemStop [0]{}%
\providecommand \bibitemNoStop [0]{.\EOS\space}%
\providecommand \EOS [0]{\spacefactor3000\relax}%
\providecommand \BibitemShut  [1]{\csname bibitem#1\endcsname}%
\let\auto@bib@innerbib\@empty
\bibitem [{\citenamefont {{Huang}}(1987)}]{huang_87}%
  \BibitemOpen
  \bibfield  {author} {\bibinfo {author} {\bibfnamefont {K.}~\bibnamefont {{Huang}}},\ }\href {https://ui.adsabs.harvard.edu/abs/1987stme.book.....H} {\emph {\bibinfo {title} {Statistical Mechanics, 2nd Edition}}}\ (\bibinfo  {publisher} {John Wiley \& Sons},\ \bibinfo {year} {1987})\BibitemShut {NoStop}%
\bibitem [{\citenamefont {{Landau}}\ and\ \citenamefont {{Lifshitz}}(1959)}]{landau_lifshitz_59}%
  \BibitemOpen
  \bibfield  {author} {\bibinfo {author} {\bibfnamefont {L.~D.}\ \bibnamefont {{Landau}}}\ and\ \bibinfo {author} {\bibfnamefont {E.~M.}\ \bibnamefont {{Lifshitz}}},\ }\href {https://ui.adsabs.harvard.edu/abs/1959flme.book.....L} {\emph {\bibinfo {title} {Fluid mechanics}}}\ (\bibinfo  {publisher} {Pergamon Press},\ \bibinfo {year} {1959})\BibitemShut {NoStop}%
\bibitem [{\citenamefont {Narozhny}(2023)}]{narozhny_23}%
  \BibitemOpen
  \bibfield  {author} {\bibinfo {author} {\bibfnamefont {B.~N.}\ \bibnamefont {Narozhny}},\ }\bibfield  {title} {\bibinfo {title} {Hydrodynamic approach to many-body systems: Exact conservation laws},\ }\href {https://doi.org/10.1016/j.aop.2023.169341} {\bibfield  {journal} {\bibinfo  {journal} {Ann. Phys.}\ }\textbf {\bibinfo {volume} {454}},\ \bibinfo {pages} {169341} (\bibinfo {year} {2023})}\BibitemShut {NoStop}%
\bibitem [{\citenamefont {Bjorken}(1983)}]{bjorken_83}%
  \BibitemOpen
  \bibfield  {author} {\bibinfo {author} {\bibfnamefont {J.~D.}\ \bibnamefont {Bjorken}},\ }\bibfield  {title} {\bibinfo {title} {Highly relativistic nucleus-nucleus collisions: The central rapidity region},\ }\href {https://doi.org/10.1103/PhysRevD.27.140} {\bibfield  {journal} {\bibinfo  {journal} {Phys. Rev. D}\ }\textbf {\bibinfo {volume} {27}},\ \bibinfo {pages} {140} (\bibinfo {year} {1983})}\BibitemShut {NoStop}%
\bibitem [{\citenamefont {Heller}\ \emph {et~al.}(2018)\citenamefont {Heller}, \citenamefont {Kurkela}, \citenamefont {Spali\ifmmode~\acute{n}\else \'{n}\fi{}ski},\ and\ \citenamefont {Svensson}}]{heller_kurkela_18}%
  \BibitemOpen
  \bibfield  {author} {\bibinfo {author} {\bibfnamefont {M.~P.}\ \bibnamefont {Heller}}, \bibinfo {author} {\bibfnamefont {A.}~\bibnamefont {Kurkela}}, \bibinfo {author} {\bibfnamefont {M.}~\bibnamefont {Spali\ifmmode~\acute{n}\else \'{n}\fi{}ski}},\ and\ \bibinfo {author} {\bibfnamefont {V.}~\bibnamefont {Svensson}},\ }\bibfield  {title} {\bibinfo {title} {Hydrodynamization in kinetic theory: Transient modes and the gradient expansion},\ }\href {https://doi.org/10.1103/PhysRevD.97.091503} {\bibfield  {journal} {\bibinfo  {journal} {Phys. Rev. D}\ }\textbf {\bibinfo {volume} {97}},\ \bibinfo {pages} {091503(R)} (\bibinfo {year} {2018})}\BibitemShut {NoStop}%
\bibitem [{\citenamefont {Florkowski}\ \emph {et~al.}(2018)\citenamefont {Florkowski}, \citenamefont {Heller},\ and\ \citenamefont {Spali\'{n}ski}}]{florkowski_heller_18}%
  \BibitemOpen
  \bibfield  {author} {\bibinfo {author} {\bibfnamefont {W.}~\bibnamefont {Florkowski}}, \bibinfo {author} {\bibfnamefont {M.~P.}\ \bibnamefont {Heller}},\ and\ \bibinfo {author} {\bibfnamefont {M.}~\bibnamefont {Spali\'{n}ski}},\ }\bibfield  {title} {\bibinfo {title} {New theories of relativistic hydrodynamics in the {LHC} era},\ }\href {https://doi.org/10.1088/1361-6633/aaa091} {\bibfield  {journal} {\bibinfo  {journal} {Rep. Prog. Phys.}\ }\textbf {\bibinfo {volume} {81}},\ \bibinfo {pages} {046001} (\bibinfo {year} {2018})}\BibitemShut {NoStop}%
\bibitem [{\citenamefont {Kurkela}\ \emph {et~al.}(2020)\citenamefont {Kurkela}, \citenamefont {Taghavi}, \citenamefont {Wiedemann},\ and\ \citenamefont {Wu}}]{kurkela_taghavi_20}%
  \BibitemOpen
  \bibfield  {author} {\bibinfo {author} {\bibfnamefont {A.}~\bibnamefont {Kurkela}}, \bibinfo {author} {\bibfnamefont {S.~F.}\ \bibnamefont {Taghavi}}, \bibinfo {author} {\bibfnamefont {U.~A.}\ \bibnamefont {Wiedemann}},\ and\ \bibinfo {author} {\bibfnamefont {B.}~\bibnamefont {Wu}},\ }\bibfield  {title} {\bibinfo {title} {Hydrodynamization in systems with detailed transverse profiles},\ }\href {https://doi.org/10.1016/j.physletb.2020.135901} {\bibfield  {journal} {\bibinfo  {journal} {Phys. Lett. B}\ }\textbf {\bibinfo {volume} {811}},\ \bibinfo {pages} {135901} (\bibinfo {year} {2020})}\BibitemShut {NoStop}%
\bibitem [{\citenamefont {Shen}\ and\ \citenamefont {Yan}(2020)}]{shen_yan_20}%
  \BibitemOpen
  \bibfield  {author} {\bibinfo {author} {\bibfnamefont {C.}~\bibnamefont {Shen}}\ and\ \bibinfo {author} {\bibfnamefont {L.}~\bibnamefont {Yan}},\ }\bibfield  {title} {\bibinfo {title} {Recent development of hydrodynamic modeling in heavy-ion collisions},\ }\href {https://doi.org/10.1007/s41365-020-00829-z} {\bibfield  {journal} {\bibinfo  {journal} {Nucl. Sci. Tech.}\ }\textbf {\bibinfo {volume} {31}},\ \bibinfo {pages} {122} (\bibinfo {year} {2020})}\BibitemShut {NoStop}%
\bibitem [{\citenamefont {Strickland}(2024)}]{strickland_24}%
  \BibitemOpen
  \bibfield  {author} {\bibinfo {author} {\bibfnamefont {M.}~\bibnamefont {Strickland}},\ }\bibfield  {title} {\bibinfo {title} {Hydrodynamization and resummed viscous hydrodynamics},\ }\href {https://arxiv.org/abs/2402.09571} {\bibfield  {journal} {\bibinfo  {journal} {arXiv:2402.09571}\ } (\bibinfo {year} {2024})}\BibitemShut {NoStop}%
\bibitem [{\citenamefont {de~Brito}\ and\ \citenamefont {Denicol}(2025)}]{debrito_denicol_25}%
  \BibitemOpen
  \bibfield  {author} {\bibinfo {author} {\bibfnamefont {C.~V.~P.}\ \bibnamefont {de~Brito}}\ and\ \bibinfo {author} {\bibfnamefont {G.~S.}\ \bibnamefont {Denicol}},\ }\bibfield  {title} {\bibinfo {title} {Hydrodynamization and thermalization in heavy-ion collisions: a kinetic theory perspective},\ }\href {https://arxiv.org/abs/2510.14547} {\bibfield  {journal} {\bibinfo  {journal} {arXiv:2510.14547}\ } (\bibinfo {year} {2025})}\BibitemShut {NoStop}%
\bibitem [{\citenamefont {Lu}\ and\ \citenamefont {Shi}(2025)}]{lu_shi_25}%
  \BibitemOpen
  \bibfield  {author} {\bibinfo {author} {\bibfnamefont {X.}~\bibnamefont {Lu}}\ and\ \bibinfo {author} {\bibfnamefont {S.}~\bibnamefont {Shi}},\ }\bibfield  {title} {\bibinfo {title} {Decoupling hydrodynamization from thermalization via nonlinear {B}oltzmann equation},\ }\href {https://arxiv.org/abs/2509.23978} {\bibfield  {journal} {\bibinfo  {journal} {arXiv:2509.23978}\ } (\bibinfo {year} {2025})}\BibitemShut {NoStop}%
\bibitem [{\citenamefont {Abdi}\ and\ \citenamefont {Nonaka}(2026)}]{abdi_nonaka_26}%
  \BibitemOpen
  \bibfield  {author} {\bibinfo {author} {\bibfnamefont {C.}~\bibnamefont {Abdi}}\ and\ \bibinfo {author} {\bibfnamefont {C.}~\bibnamefont {Nonaka}},\ }\bibfield  {title} {\bibinfo {title} {Chemical equilibration and thermalization of quark-gluon plasma in a parton cascade model with 2-to-3 quark interactions},\ }\href {https://arxiv.org/abs/2606.03058} {\bibfield  {journal} {\bibinfo  {journal} {arXiv:2606.03058}\ } (\bibinfo {year} {2026})}\BibitemShut {NoStop}%
\bibitem [{\citenamefont {Rajagopal}\ \emph {et~al.}(2026)\citenamefont {Rajagopal}, \citenamefont {Scheihing-Hitschfeld},\ and\ \citenamefont {Steinhorst}}]{rajagopal_scheihing-hitschfeld_26}%
  \BibitemOpen
  \bibfield  {author} {\bibinfo {author} {\bibfnamefont {K.}~\bibnamefont {Rajagopal}}, \bibinfo {author} {\bibfnamefont {B.}~\bibnamefont {Scheihing-Hitschfeld}},\ and\ \bibinfo {author} {\bibfnamefont {R.}~\bibnamefont {Steinhorst}},\ }\bibfield  {title} {\bibinfo {title} {Attractors without scaling: adiabatic hydrodynamization with and without inelastic scattering},\ }\href {https://doi.org/10.1007/JHEP03(2026)003} {\bibfield  {journal} {\bibinfo  {journal} {J. High Energy Phys.}\ }\textbf {\bibinfo {volume} {2026}},\ \bibinfo {pages} {3}}\BibitemShut {NoStop}%
\bibitem [{\citenamefont {Le}\ \emph {et~al.}(2023)\citenamefont {Le}, \citenamefont {Zhang}, \citenamefont {Gopalakrishnan}, \citenamefont {Rigol},\ and\ \citenamefont {Weiss}}]{le_zhang_23}%
  \BibitemOpen
  \bibfield  {author} {\bibinfo {author} {\bibfnamefont {Y.}~\bibnamefont {Le}}, \bibinfo {author} {\bibfnamefont {Y.}~\bibnamefont {Zhang}}, \bibinfo {author} {\bibfnamefont {S.}~\bibnamefont {Gopalakrishnan}}, \bibinfo {author} {\bibfnamefont {M.}~\bibnamefont {Rigol}},\ and\ \bibinfo {author} {\bibfnamefont {D.~S.}\ \bibnamefont {Weiss}},\ }\bibfield  {title} {\bibinfo {title} {Observation of hydrodynamization and local prethermalization in {1D Bose} gases},\ }\href {https://doi.org/10.1038/s41586-023-05979-9} {\bibfield  {journal} {\bibinfo  {journal} {Nature}\ }\textbf {\bibinfo {volume} {618}},\ \bibinfo {pages} {494} (\bibinfo {year} {2023})}\BibitemShut {NoStop}%
\bibitem [{\citenamefont {Zhang}\ \emph {et~al.}(2025)\citenamefont {Zhang}, \citenamefont {Le}, \citenamefont {Weiss},\ and\ \citenamefont {Rigol}}]{zhang_le_25}%
  \BibitemOpen
  \bibfield  {author} {\bibinfo {author} {\bibfnamefont {Y.}~\bibnamefont {Zhang}}, \bibinfo {author} {\bibfnamefont {Y.}~\bibnamefont {Le}}, \bibinfo {author} {\bibfnamefont {D.~S.}\ \bibnamefont {Weiss}},\ and\ \bibinfo {author} {\bibfnamefont {M.}~\bibnamefont {Rigol}},\ }\bibfield  {title} {\bibinfo {title} {Timescales and necessary conditions for hydrodynamization in one-dimensional {Bose} gases},\ }\href {https://doi.org/10.1103/PhysRevA.111.053306} {\bibfield  {journal} {\bibinfo  {journal} {Phys. Rev. A}\ }\textbf {\bibinfo {volume} {111}},\ \bibinfo {pages} {053306} (\bibinfo {year} {2025})}\BibitemShut {NoStop}%
\bibitem [{\citenamefont {Rigol}\ \emph {et~al.}(2006)\citenamefont {Rigol}, \citenamefont {Muramatsu},\ and\ \citenamefont {Olshanii}}]{rigol_muramatsu_06}%
  \BibitemOpen
  \bibfield  {author} {\bibinfo {author} {\bibfnamefont {M.}~\bibnamefont {Rigol}}, \bibinfo {author} {\bibfnamefont {A.}~\bibnamefont {Muramatsu}},\ and\ \bibinfo {author} {\bibfnamefont {M.}~\bibnamefont {Olshanii}},\ }\bibfield  {title} {\bibinfo {title} {Hard-core bosons on optical superlattices: {D}ynamics and relaxation in the superfluid and insulating regimes},\ }\href {https://doi.org/10.1103/PhysRevA.74.053616} {\bibfield  {journal} {\bibinfo  {journal} {Phys. Rev. A}\ }\textbf {\bibinfo {volume} {74}},\ \bibinfo {pages} {053616} (\bibinfo {year} {2006})}\BibitemShut {NoStop}%
\bibitem [{\citenamefont {van~den Berg}\ \emph {et~al.}(2016)\citenamefont {van~den Berg}, \citenamefont {Wouters}, \citenamefont {Elie\"ns}, \citenamefont {De~Nardis}, \citenamefont {Konik},\ and\ \citenamefont {Caux}}]{caux_bragg_16}%
  \BibitemOpen
  \bibfield  {author} {\bibinfo {author} {\bibfnamefont {R.}~\bibnamefont {van~den Berg}}, \bibinfo {author} {\bibfnamefont {B.}~\bibnamefont {Wouters}}, \bibinfo {author} {\bibfnamefont {S.}~\bibnamefont {Elie\"ns}}, \bibinfo {author} {\bibfnamefont {J.}~\bibnamefont {De~Nardis}}, \bibinfo {author} {\bibfnamefont {R.~M.}\ \bibnamefont {Konik}},\ and\ \bibinfo {author} {\bibfnamefont {J.-S.}\ \bibnamefont {Caux}},\ }\bibfield  {title} {\bibinfo {title} {Separation of time scales in a quantum {N}ewton's cradle},\ }\href {https://doi.org/10.1103/PhysRevLett.116.225302} {\bibfield  {journal} {\bibinfo  {journal} {Phys. Rev. Lett.}\ }\textbf {\bibinfo {volume} {116}},\ \bibinfo {pages} {225302} (\bibinfo {year} {2016})}\BibitemShut {NoStop}%
\bibitem [{\citenamefont {Lieb}\ and\ \citenamefont {Liniger}(1963)}]{lieb_liniger_63}%
  \BibitemOpen
  \bibfield  {author} {\bibinfo {author} {\bibfnamefont {E.~H.}\ \bibnamefont {Lieb}}\ and\ \bibinfo {author} {\bibfnamefont {W.}~\bibnamefont {Liniger}},\ }\bibfield  {title} {\bibinfo {title} {Exact analysis of an interacting {Bose} gas. {I. The} general solution and the ground state},\ }\href {https://doi.org/10.1103/PhysRev.130.1605} {\bibfield  {journal} {\bibinfo  {journal} {Phys. Rev.}\ }\textbf {\bibinfo {volume} {130}},\ \bibinfo {pages} {1605} (\bibinfo {year} {1963})}\BibitemShut {NoStop}%
\bibitem [{\citenamefont {Yang}\ and\ \citenamefont {Yang}(1969)}]{yang_yang_69}%
  \BibitemOpen
  \bibfield  {author} {\bibinfo {author} {\bibfnamefont {C.~N.}\ \bibnamefont {Yang}}\ and\ \bibinfo {author} {\bibfnamefont {C.~P.}\ \bibnamefont {Yang}},\ }\bibfield  {title} {\bibinfo {title} {Thermodynamics of a one-dimensional system of bosons with repulsive delta-function interaction},\ }\href {https://doi.org/10.1063/1.1664947} {\bibfield  {journal} {\bibinfo  {journal} {J. Math. Phys.}\ }\textbf {\bibinfo {volume} {10}},\ \bibinfo {pages} {1115} (\bibinfo {year} {1969})}\BibitemShut {NoStop}%
\bibitem [{\citenamefont {Cazalilla}\ \emph {et~al.}(2011)\citenamefont {Cazalilla}, \citenamefont {Citro}, \citenamefont {Giamarchi}, \citenamefont {Orignac},\ and\ \citenamefont {Rigol}}]{cazalilla_citro_11}%
  \BibitemOpen
  \bibfield  {author} {\bibinfo {author} {\bibfnamefont {M.~A.}\ \bibnamefont {Cazalilla}}, \bibinfo {author} {\bibfnamefont {R.}~\bibnamefont {Citro}}, \bibinfo {author} {\bibfnamefont {T.}~\bibnamefont {Giamarchi}}, \bibinfo {author} {\bibfnamefont {E.}~\bibnamefont {Orignac}},\ and\ \bibinfo {author} {\bibfnamefont {M.}~\bibnamefont {Rigol}},\ }\bibfield  {title} {\bibinfo {title} {One dimensional bosons: From condensed matter systems to ultracold gases},\ }\href {https://doi.org/10.1103/RevModPhys.83.1405} {\bibfield  {journal} {\bibinfo  {journal} {Rev. Mod. Phys.}\ }\textbf {\bibinfo {volume} {83}},\ \bibinfo {pages} {1405} (\bibinfo {year} {2011})}\BibitemShut {NoStop}%
\bibitem [{\citenamefont {Rigol}\ \emph {et~al.}(2007)\citenamefont {Rigol}, \citenamefont {Dunjko}, \citenamefont {Yurovsky},\ and\ \citenamefont {Olshanii}}]{rigol_dunjko_07}%
  \BibitemOpen
  \bibfield  {author} {\bibinfo {author} {\bibfnamefont {M.}~\bibnamefont {Rigol}}, \bibinfo {author} {\bibfnamefont {V.}~\bibnamefont {Dunjko}}, \bibinfo {author} {\bibfnamefont {V.}~\bibnamefont {Yurovsky}},\ and\ \bibinfo {author} {\bibfnamefont {M.}~\bibnamefont {Olshanii}},\ }\bibfield  {title} {\bibinfo {title} {Relaxation in a completely integrable many-body quantum system: An ab initio study of the dynamics of the highly excited states of {1D} lattice hard-core bosons},\ }\href {https://doi.org/10.1103/PhysRevLett.98.050405} {\bibfield  {journal} {\bibinfo  {journal} {Phys. Rev. Lett.}\ }\textbf {\bibinfo {volume} {98}},\ \bibinfo {pages} {050405} (\bibinfo {year} {2007})}\BibitemShut {NoStop}%
\bibitem [{\citenamefont {Wouters}\ \emph {et~al.}(2014)\citenamefont {Wouters}, \citenamefont {De~Nardis}, \citenamefont {Brockmann}, \citenamefont {Fioretto}, \citenamefont {Rigol},\ and\ \citenamefont {Caux}}]{wouters_denardis_14}%
  \BibitemOpen
  \bibfield  {author} {\bibinfo {author} {\bibfnamefont {B.}~\bibnamefont {Wouters}}, \bibinfo {author} {\bibfnamefont {J.}~\bibnamefont {De~Nardis}}, \bibinfo {author} {\bibfnamefont {M.}~\bibnamefont {Brockmann}}, \bibinfo {author} {\bibfnamefont {D.}~\bibnamefont {Fioretto}}, \bibinfo {author} {\bibfnamefont {M.}~\bibnamefont {Rigol}},\ and\ \bibinfo {author} {\bibfnamefont {J.-S.}\ \bibnamefont {Caux}},\ }\bibfield  {title} {\bibinfo {title} {Quenching the anisotropic {H}eisenberg chain: Exact solution and generalized {G}ibbs ensemble predictions},\ }\href {https://doi.org/10.1103/PhysRevLett.113.117202} {\bibfield  {journal} {\bibinfo  {journal} {Phys. Rev. Lett.}\ }\textbf {\bibinfo {volume} {113}},\ \bibinfo {pages} {117202} (\bibinfo {year} {2014})}\BibitemShut {NoStop}%
\bibitem [{\citenamefont {Pozsgay}\ \emph {et~al.}(2014)\citenamefont {Pozsgay}, \citenamefont {Mesty\'an}, \citenamefont {Werner}, \citenamefont {Kormos}, \citenamefont {Zar\'and},\ and\ \citenamefont {Tak\'acs}}]{pozsgay_mestyan_14}%
  \BibitemOpen
  \bibfield  {author} {\bibinfo {author} {\bibfnamefont {B.}~\bibnamefont {Pozsgay}}, \bibinfo {author} {\bibfnamefont {M.}~\bibnamefont {Mesty\'an}}, \bibinfo {author} {\bibfnamefont {M.~A.}\ \bibnamefont {Werner}}, \bibinfo {author} {\bibfnamefont {M.}~\bibnamefont {Kormos}}, \bibinfo {author} {\bibfnamefont {G.}~\bibnamefont {Zar\'and}},\ and\ \bibinfo {author} {\bibfnamefont {G.}~\bibnamefont {Tak\'acs}},\ }\bibfield  {title} {\bibinfo {title} {Correlations after quantum quenches in the {$XXZ$} spin chain: Failure of the generalized {G}ibbs ensemble},\ }\href {https://doi.org/10.1103/PhysRevLett.113.117203} {\bibfield  {journal} {\bibinfo  {journal} {Phys. Rev. Lett.}\ }\textbf {\bibinfo {volume} {113}},\ \bibinfo {pages} {117203} (\bibinfo {year} {2014})}\BibitemShut {NoStop}%
\bibitem [{\citenamefont {Ilievski}\ \emph {et~al.}(2015)\citenamefont {Ilievski}, \citenamefont {De~Nardis}, \citenamefont {Wouters}, \citenamefont {Caux}, \citenamefont {Essler},\ and\ \citenamefont {Prosen}}]{ilievski_denardis_15}%
  \BibitemOpen
  \bibfield  {author} {\bibinfo {author} {\bibfnamefont {E.}~\bibnamefont {Ilievski}}, \bibinfo {author} {\bibfnamefont {J.}~\bibnamefont {De~Nardis}}, \bibinfo {author} {\bibfnamefont {B.}~\bibnamefont {Wouters}}, \bibinfo {author} {\bibfnamefont {J.-S.}\ \bibnamefont {Caux}}, \bibinfo {author} {\bibfnamefont {F.~H.~L.}\ \bibnamefont {Essler}},\ and\ \bibinfo {author} {\bibfnamefont {T.}~\bibnamefont {Prosen}},\ }\bibfield  {title} {\bibinfo {title} {Complete generalized {G}ibbs ensembles in an interacting theory},\ }\href {https://doi.org/10.1103/PhysRevLett.115.157201} {\bibfield  {journal} {\bibinfo  {journal} {Phys. Rev. Lett.}\ }\textbf {\bibinfo {volume} {115}},\ \bibinfo {pages} {157201} (\bibinfo {year} {2015})}\BibitemShut {NoStop}%
\bibitem [{\citenamefont {Vidmar}\ and\ \citenamefont {Rigol}(2016)}]{vidmar_rigol_16}%
  \BibitemOpen
  \bibfield  {author} {\bibinfo {author} {\bibfnamefont {L.}~\bibnamefont {Vidmar}}\ and\ \bibinfo {author} {\bibfnamefont {M.}~\bibnamefont {Rigol}},\ }\bibfield  {title} {\bibinfo {title} {Generalized {G}ibbs ensemble in integrable lattice models},\ }\href {https://doi.org/10.1088/1742-5468/2016/06/064007} {\bibfield  {journal} {\bibinfo  {journal} {J. Stat. Mech.}\ }\textbf {\bibinfo {volume} {2016}},\ \bibinfo {pages} {064007} (\bibinfo {year} {2016})}\BibitemShut {NoStop}%
\bibitem [{\citenamefont {Wilson}\ \emph {et~al.}(2020)\citenamefont {Wilson}, \citenamefont {Malvania}, \citenamefont {Le}, \citenamefont {Zhang}, \citenamefont {Rigol},\ and\ \citenamefont {Weiss}}]{wilson_malvania_20}%
  \BibitemOpen
  \bibfield  {author} {\bibinfo {author} {\bibfnamefont {J.~M.}\ \bibnamefont {Wilson}}, \bibinfo {author} {\bibfnamefont {N.}~\bibnamefont {Malvania}}, \bibinfo {author} {\bibfnamefont {Y.}~\bibnamefont {Le}}, \bibinfo {author} {\bibfnamefont {Y.}~\bibnamefont {Zhang}}, \bibinfo {author} {\bibfnamefont {M.}~\bibnamefont {Rigol}},\ and\ \bibinfo {author} {\bibfnamefont {D.~S.}\ \bibnamefont {Weiss}},\ }\bibfield  {title} {\bibinfo {title} {Observation of dynamical fermionization},\ }\href {https://doi.org/10.1126/science.aaz0242} {\bibfield  {journal} {\bibinfo  {journal} {Science}\ }\textbf {\bibinfo {volume} {367}},\ \bibinfo {pages} {1461} (\bibinfo {year} {2020})}\BibitemShut {NoStop}%
\bibitem [{\citenamefont {Li}\ \emph {et~al.}(2023)\citenamefont {Li}, \citenamefont {Zhang}, \citenamefont {Yang}, \citenamefont {Lin}, \citenamefont {Gopalakrishnan}, \citenamefont {Rigol},\ and\ \citenamefont {Lev}}]{li_zhang_23}%
  \BibitemOpen
  \bibfield  {author} {\bibinfo {author} {\bibfnamefont {K.-Y.}\ \bibnamefont {Li}}, \bibinfo {author} {\bibfnamefont {Y.}~\bibnamefont {Zhang}}, \bibinfo {author} {\bibfnamefont {K.}~\bibnamefont {Yang}}, \bibinfo {author} {\bibfnamefont {K.-Y.}\ \bibnamefont {Lin}}, \bibinfo {author} {\bibfnamefont {S.}~\bibnamefont {Gopalakrishnan}}, \bibinfo {author} {\bibfnamefont {M.}~\bibnamefont {Rigol}},\ and\ \bibinfo {author} {\bibfnamefont {B.~L.}\ \bibnamefont {Lev}},\ }\bibfield  {title} {\bibinfo {title} {Rapidity and momentum distributions of one-dimensional dipolar quantum gases},\ }\href {https://doi.org/10.1103/PhysRevA.107.L061302} {\bibfield  {journal} {\bibinfo  {journal} {Phys. Rev. A}\ }\textbf {\bibinfo {volume} {107}},\ \bibinfo {pages} {L061302} (\bibinfo {year} {2023})}\BibitemShut {NoStop}%
\bibitem [{\citenamefont {Dubois}\ \emph {et~al.}(2024)\citenamefont {Dubois}, \citenamefont {Th\'em\`eze}, \citenamefont {Nogrette}, \citenamefont {Dubail},\ and\ \citenamefont {Bouchoule}}]{Dubois2024Probing}%
  \BibitemOpen
  \bibfield  {author} {\bibinfo {author} {\bibfnamefont {L.}~\bibnamefont {Dubois}}, \bibinfo {author} {\bibfnamefont {G.}~\bibnamefont {Th\'em\`eze}}, \bibinfo {author} {\bibfnamefont {F.}~\bibnamefont {Nogrette}}, \bibinfo {author} {\bibfnamefont {J.}~\bibnamefont {Dubail}},\ and\ \bibinfo {author} {\bibfnamefont {I.}~\bibnamefont {Bouchoule}},\ }\bibfield  {title} {\bibinfo {title} {Probing the local rapidity distribution of a one-dimensional {B}ose gas},\ }\href {https://doi.org/10.1103/PhysRevLett.133.113402} {\bibfield  {journal} {\bibinfo  {journal} {Phys. Rev. Lett.}\ }\textbf {\bibinfo {volume} {133}},\ \bibinfo {pages} {113402} (\bibinfo {year} {2024})}\BibitemShut {NoStop}%
\bibitem [{\citenamefont {Dhar}\ \emph {et~al.}(2025)\citenamefont {Dhar}, \citenamefont {Wang}, \citenamefont {Horvath}, \citenamefont {Vashisht}, \citenamefont {Zeng}, \citenamefont {Zvonarev}, \citenamefont {Goldman}, \citenamefont {Guo}, \citenamefont {Landini},\ and\ \citenamefont {N{\"a}gerl}}]{dhar2025anyonization}%
  \BibitemOpen
  \bibfield  {author} {\bibinfo {author} {\bibfnamefont {S.}~\bibnamefont {Dhar}}, \bibinfo {author} {\bibfnamefont {B.}~\bibnamefont {Wang}}, \bibinfo {author} {\bibfnamefont {M.}~\bibnamefont {Horvath}}, \bibinfo {author} {\bibfnamefont {A.}~\bibnamefont {Vashisht}}, \bibinfo {author} {\bibfnamefont {Y.}~\bibnamefont {Zeng}}, \bibinfo {author} {\bibfnamefont {M.~B.}\ \bibnamefont {Zvonarev}}, \bibinfo {author} {\bibfnamefont {N.}~\bibnamefont {Goldman}}, \bibinfo {author} {\bibfnamefont {Y.}~\bibnamefont {Guo}}, \bibinfo {author} {\bibfnamefont {M.}~\bibnamefont {Landini}},\ and\ \bibinfo {author} {\bibfnamefont {H.-C.}\ \bibnamefont {N{\"a}gerl}},\ }\bibfield  {title} {\bibinfo {title} {Observing anyonization of bosons in a quantum gas},\ }\href {https://doi.org/10.1038/s41586-025-09016-9} {\bibfield  {journal} {\bibinfo  {journal} {Nature}\ }\textbf {\bibinfo {volume} {642}},\ \bibinfo {pages} {53} (\bibinfo {year} {2025})}\BibitemShut {NoStop}%
\bibitem [{\citenamefont {Yang}\ \emph {et~al.}(2024)\citenamefont {Yang}, \citenamefont {Zhang}, \citenamefont {Li}, \citenamefont {Lin}, \citenamefont {Gopalakrishnan}, \citenamefont {Rigol},\ and\ \citenamefont {Lev}}]{Yang2024Phantom}%
  \BibitemOpen
  \bibfield  {author} {\bibinfo {author} {\bibfnamefont {K.}~\bibnamefont {Yang}}, \bibinfo {author} {\bibfnamefont {Y.}~\bibnamefont {Zhang}}, \bibinfo {author} {\bibfnamefont {K.-Y.}\ \bibnamefont {Li}}, \bibinfo {author} {\bibfnamefont {K.-Y.}\ \bibnamefont {Lin}}, \bibinfo {author} {\bibfnamefont {S.}~\bibnamefont {Gopalakrishnan}}, \bibinfo {author} {\bibfnamefont {M.}~\bibnamefont {Rigol}},\ and\ \bibinfo {author} {\bibfnamefont {B.~L.}\ \bibnamefont {Lev}},\ }\bibfield  {title} {\bibinfo {title} {Phantom energy in the nonlinear response of a quantum many-body scar state},\ }\href {https://doi.org/10.1126/science.adk8978} {\bibfield  {journal} {\bibinfo  {journal} {Science}\ }\textbf {\bibinfo {volume} {385}},\ \bibinfo {pages} {1063} (\bibinfo {year} {2024})}\BibitemShut {NoStop}%
\bibitem [{\citenamefont {Horvath}\ \emph {et~al.}()\citenamefont {Horvath}, \citenamefont {Bastianello}, \citenamefont {Dhar}, \citenamefont {Koch}, \citenamefont {Guo}, \citenamefont {Caux}, \citenamefont {Landini},\ and\ \citenamefont {Nägerl}}]{horvath2025observing}%
  \BibitemOpen
  \bibfield  {author} {\bibinfo {author} {\bibfnamefont {M.}~\bibnamefont {Horvath}}, \bibinfo {author} {\bibfnamefont {A.}~\bibnamefont {Bastianello}}, \bibinfo {author} {\bibfnamefont {S.}~\bibnamefont {Dhar}}, \bibinfo {author} {\bibfnamefont {R.}~\bibnamefont {Koch}}, \bibinfo {author} {\bibfnamefont {Y.}~\bibnamefont {Guo}}, \bibinfo {author} {\bibfnamefont {J.-S.}\ \bibnamefont {Caux}}, \bibinfo {author} {\bibfnamefont {M.}~\bibnamefont {Landini}},\ and\ \bibinfo {author} {\bibfnamefont {H.-C.}\ \bibnamefont {Nägerl}},\ }\href {https://arxiv.org/abs/2505.10550} {\bibinfo {title} {Observing {B}ethe strings in an attractive {B}ose gas far from equilibrium}},\ \Eprint {https://arxiv.org/abs/2505.10550} {arXiv:2505.10550} \BibitemShut {NoStop}%
\bibitem [{\citenamefont {Bastianello}\ \emph {et~al.}(2026)\citenamefont {Bastianello}, \citenamefont {Zeng}, \citenamefont {Dhar}, \citenamefont {Wang}, \citenamefont {Yu}, \citenamefont {Horvath}, \citenamefont {Astrakharchik}, \citenamefont {Guo}, \citenamefont {N\"agerl},\ and\ \citenamefont {Landini}}]{bastianello_zeng_26}%
  \BibitemOpen
  \bibfield  {author} {\bibinfo {author} {\bibfnamefont {A.}~\bibnamefont {Bastianello}}, \bibinfo {author} {\bibfnamefont {Y.}~\bibnamefont {Zeng}}, \bibinfo {author} {\bibfnamefont {S.}~\bibnamefont {Dhar}}, \bibinfo {author} {\bibfnamefont {Z.}~\bibnamefont {Wang}}, \bibinfo {author} {\bibfnamefont {X.}~\bibnamefont {Yu}}, \bibinfo {author} {\bibfnamefont {M.}~\bibnamefont {Horvath}}, \bibinfo {author} {\bibfnamefont {G.~E.}\ \bibnamefont {Astrakharchik}}, \bibinfo {author} {\bibfnamefont {Y.}~\bibnamefont {Guo}}, \bibinfo {author} {\bibfnamefont {H.-C.}\ \bibnamefont {N\"agerl}},\ and\ \bibinfo {author} {\bibfnamefont {M.}~\bibnamefont {Landini}},\ }\bibfield  {title} {\bibinfo {title} {Exotic critical states as fractional {F}ermi seas in the one-dimensional {B}ose gas},\ }\href {https://doi.org/10.1103/j3s5-gjpf} {\bibfield  {journal} {\bibinfo  {journal} {Phys. Rev. Lett.}\ }\textbf {\bibinfo {volume} {136}},\ \bibinfo {pages} {230402} (\bibinfo {year} {2026})}\BibitemShut {NoStop}%
\bibitem [{\citenamefont {Castro-Alvaredo}\ \emph {et~al.}(2016)\citenamefont {Castro-Alvaredo}, \citenamefont {Doyon},\ and\ \citenamefont {Yoshimura}}]{castro2016emergent}%
  \BibitemOpen
  \bibfield  {author} {\bibinfo {author} {\bibfnamefont {O.~A.}\ \bibnamefont {Castro-Alvaredo}}, \bibinfo {author} {\bibfnamefont {B.}~\bibnamefont {Doyon}},\ and\ \bibinfo {author} {\bibfnamefont {T.}~\bibnamefont {Yoshimura}},\ }\bibfield  {title} {\bibinfo {title} {Emergent hydrodynamics in integrable quantum systems out of equilibrium},\ }\href {https://doi.org/10.1103/PhysRevX.6.041065} {\bibfield  {journal} {\bibinfo  {journal} {Phys. Rev. X}\ }\textbf {\bibinfo {volume} {6}},\ \bibinfo {pages} {041065} (\bibinfo {year} {2016})}\BibitemShut {NoStop}%
\bibitem [{\citenamefont {Bertini}\ \emph {et~al.}(2016)\citenamefont {Bertini}, \citenamefont {Collura}, \citenamefont {De~Nardis},\ and\ \citenamefont {Fagotti}}]{bertini_collura_16}%
  \BibitemOpen
  \bibfield  {author} {\bibinfo {author} {\bibfnamefont {B.}~\bibnamefont {Bertini}}, \bibinfo {author} {\bibfnamefont {M.}~\bibnamefont {Collura}}, \bibinfo {author} {\bibfnamefont {J.}~\bibnamefont {De~Nardis}},\ and\ \bibinfo {author} {\bibfnamefont {M.}~\bibnamefont {Fagotti}},\ }\bibfield  {title} {\bibinfo {title} {Transport in out-of-equilibrium {XXZ} chains: {E}xact profiles of charges and currents},\ }\href {https://doi.org/10.1103/PhysRevLett.117.207201} {\bibfield  {journal} {\bibinfo  {journal} {Phys. Rev. Lett.}\ }\textbf {\bibinfo {volume} {117}},\ \bibinfo {pages} {207201} (\bibinfo {year} {2016})}\BibitemShut {NoStop}%
\bibitem [{\citenamefont {Alba}\ \emph {et~al.}(2021)\citenamefont {Alba}, \citenamefont {Bertini}, \citenamefont {Fagotti}, \citenamefont {Piroli},\ and\ \citenamefont {Ruggiero}}]{alba_bertini_21}%
  \BibitemOpen
  \bibfield  {author} {\bibinfo {author} {\bibfnamefont {V.}~\bibnamefont {Alba}}, \bibinfo {author} {\bibfnamefont {B.}~\bibnamefont {Bertini}}, \bibinfo {author} {\bibfnamefont {M.}~\bibnamefont {Fagotti}}, \bibinfo {author} {\bibfnamefont {L.}~\bibnamefont {Piroli}},\ and\ \bibinfo {author} {\bibfnamefont {P.}~\bibnamefont {Ruggiero}},\ }\bibfield  {title} {\bibinfo {title} {Generalized-hydrodynamic approach to inhomogeneous quenches: correlations, entanglement and quantum effects},\ }\href {https://doi.org/10.1088/1742-5468/ac257d} {\bibfield  {journal} {\bibinfo  {journal} {J. Stat. Mech.}\ }\textbf {\bibinfo {volume} {2021}},\ \bibinfo {pages} {114004} (\bibinfo {year} {2021})}\BibitemShut {NoStop}%
\bibitem [{\citenamefont {Doyon}(2020)}]{doyon_20}%
  \BibitemOpen
  \bibfield  {author} {\bibinfo {author} {\bibfnamefont {B.}~\bibnamefont {Doyon}},\ }\bibfield  {title} {\bibinfo {title} {Lecture notes on generalised hydrodynamics},\ }\bibfield  {journal} {\bibinfo  {journal} {SciPost Phys. Lect. Notes}\ }\href {https://doi.org/10.21468/SciPostPhysLectNotes.18} {10.21468/SciPostPhysLectNotes.18} (\bibinfo {year} {2020})\BibitemShut {NoStop}%
\bibitem [{\citenamefont {Essler}(2023)}]{essler_23}%
  \BibitemOpen
  \bibfield  {author} {\bibinfo {author} {\bibfnamefont {F.~H.}\ \bibnamefont {Essler}},\ }\bibfield  {title} {\bibinfo {title} {A short introduction to generalized hydrodynamics},\ }\href {https://doi.org/10.1016/j.physa.2022.127572} {\bibfield  {journal} {\bibinfo  {journal} {Physica A}\ }\textbf {\bibinfo {volume} {631}},\ \bibinfo {pages} {127572} (\bibinfo {year} {2023})}\BibitemShut {NoStop}%
\bibitem [{\citenamefont {Doyon}\ \emph {et~al.}(2025)\citenamefont {Doyon}, \citenamefont {Gopalakrishnan}, \citenamefont {M\o{}ller}, \citenamefont {Schmiedmayer},\ and\ \citenamefont {Vasseur}}]{doyon_gopalakrishnan_25}%
  \BibitemOpen
  \bibfield  {author} {\bibinfo {author} {\bibfnamefont {B.}~\bibnamefont {Doyon}}, \bibinfo {author} {\bibfnamefont {S.}~\bibnamefont {Gopalakrishnan}}, \bibinfo {author} {\bibfnamefont {F.}~\bibnamefont {M\o{}ller}}, \bibinfo {author} {\bibfnamefont {J.}~\bibnamefont {Schmiedmayer}},\ and\ \bibinfo {author} {\bibfnamefont {R.}~\bibnamefont {Vasseur}},\ }\bibfield  {title} {\bibinfo {title} {Generalized hydrodynamics: A perspective},\ }\href {https://doi.org/10.1103/PhysRevX.15.010501} {\bibfield  {journal} {\bibinfo  {journal} {Phys. Rev. X}\ }\textbf {\bibinfo {volume} {15}},\ \bibinfo {pages} {010501} (\bibinfo {year} {2025})}\BibitemShut {NoStop}%
\bibitem [{\citenamefont {Malvania}\ \emph {et~al.}(2021)\citenamefont {Malvania}, \citenamefont {Zhang}, \citenamefont {Le}, \citenamefont {Dubail}, \citenamefont {Rigol},\ and\ \citenamefont {Weiss}}]{malvania_zhang_21}%
  \BibitemOpen
  \bibfield  {author} {\bibinfo {author} {\bibfnamefont {N.}~\bibnamefont {Malvania}}, \bibinfo {author} {\bibfnamefont {Y.}~\bibnamefont {Zhang}}, \bibinfo {author} {\bibfnamefont {Y.}~\bibnamefont {Le}}, \bibinfo {author} {\bibfnamefont {J.}~\bibnamefont {Dubail}}, \bibinfo {author} {\bibfnamefont {M.}~\bibnamefont {Rigol}},\ and\ \bibinfo {author} {\bibfnamefont {D.~S.}\ \bibnamefont {Weiss}},\ }\bibfield  {title} {\bibinfo {title} {Generalized hydrodynamics in strongly interacting {1D Bose} gases},\ }\href {https://doi.org/10.1126/science.abf0147} {\bibfield  {journal} {\bibinfo  {journal} {Science}\ }\textbf {\bibinfo {volume} {373}},\ \bibinfo {pages} {1129} (\bibinfo {year} {2021})}\BibitemShut {NoStop}%
\bibitem [{\citenamefont {Deng}\ \emph {et~al.}(1999)\citenamefont {Deng}, \citenamefont {Hagley}, \citenamefont {Denschlag}, \citenamefont {Simsarian}, \citenamefont {Edwards}, \citenamefont {Clark}, \citenamefont {Helmerson}, \citenamefont {Rolston},\ and\ \citenamefont {Phillips}}]{deng_hagley_99}%
  \BibitemOpen
  \bibfield  {author} {\bibinfo {author} {\bibfnamefont {L.}~\bibnamefont {Deng}}, \bibinfo {author} {\bibfnamefont {E.~W.}\ \bibnamefont {Hagley}}, \bibinfo {author} {\bibfnamefont {J.}~\bibnamefont {Denschlag}}, \bibinfo {author} {\bibfnamefont {J.~E.}\ \bibnamefont {Simsarian}}, \bibinfo {author} {\bibfnamefont {M.}~\bibnamefont {Edwards}}, \bibinfo {author} {\bibfnamefont {C.~W.}\ \bibnamefont {Clark}}, \bibinfo {author} {\bibfnamefont {K.}~\bibnamefont {Helmerson}}, \bibinfo {author} {\bibfnamefont {S.~L.}\ \bibnamefont {Rolston}},\ and\ \bibinfo {author} {\bibfnamefont {W.~D.}\ \bibnamefont {Phillips}},\ }\bibfield  {title} {\bibinfo {title} {Temporal, matter-wave-dispersion talbot effect},\ }\href {https://doi.org/10.1103/PhysRevLett.83.5407} {\bibfield  {journal} {\bibinfo  {journal} {Phys. Rev. Lett.}\ }\textbf {\bibinfo {volume} {83}},\ \bibinfo {pages} {5407} (\bibinfo {year} {1999})}\BibitemShut {NoStop}%
\bibitem [{\citenamefont {Santra}\ \emph {et~al.}(2017)\citenamefont {Santra}, \citenamefont {Baals}, \citenamefont {Labouvie}, \citenamefont {Bhattacherjee}, \citenamefont {Pelster},\ and\ \citenamefont {Ott}}]{santra_baals_17}%
  \BibitemOpen
  \bibfield  {author} {\bibinfo {author} {\bibfnamefont {B.}~\bibnamefont {Santra}}, \bibinfo {author} {\bibfnamefont {C.}~\bibnamefont {Baals}}, \bibinfo {author} {\bibfnamefont {R.}~\bibnamefont {Labouvie}}, \bibinfo {author} {\bibfnamefont {A.~B.}\ \bibnamefont {Bhattacherjee}}, \bibinfo {author} {\bibfnamefont {A.}~\bibnamefont {Pelster}},\ and\ \bibinfo {author} {\bibfnamefont {H.}~\bibnamefont {Ott}},\ }\bibfield  {title} {\bibinfo {title} {Measuring finite-range phase coherence in an optical lattice using talbot interferometry},\ }\href {https://doi.org/10.1038/ncomms15601} {\bibfield  {journal} {\bibinfo  {journal} {Nat. Commun.}\ }\textbf {\bibinfo {volume} {8}},\ \bibinfo {pages} {15601} (\bibinfo {year} {2017})}\BibitemShut {NoStop}%
\bibitem [{\citenamefont {Hall}\ \emph {et~al.}(2021)\citenamefont {Hall}, \citenamefont {Yessenov}, \citenamefont {Ponomarenko},\ and\ \citenamefont {Abouraddy}}]{hall_yessenov_21}%
  \BibitemOpen
  \bibfield  {author} {\bibinfo {author} {\bibfnamefont {L.~A.}\ \bibnamefont {Hall}}, \bibinfo {author} {\bibfnamefont {M.}~\bibnamefont {Yessenov}}, \bibinfo {author} {\bibfnamefont {S.~A.}\ \bibnamefont {Ponomarenko}},\ and\ \bibinfo {author} {\bibfnamefont {A.~F.}\ \bibnamefont {Abouraddy}},\ }\bibfield  {title} {\bibinfo {title} {The space-time talbot effect},\ }\href {https://doi.org/10.1063/5.0045310} {\bibfield  {journal} {\bibinfo  {journal} {APL Photonics}\ }\textbf {\bibinfo {volume} {6}},\ \bibinfo {pages} {056105} (\bibinfo {year} {2021})}\BibitemShut {NoStop}%
\bibitem [{\citenamefont {Wu}\ \emph {et~al.}(2023)\citenamefont {Wu}, \citenamefont {Clementi}, \citenamefont {Nitiss}, \citenamefont {Hu}, \citenamefont {Lafforgue},\ and\ \citenamefont {Br{\`e}s}}]{wu_clementi_23}%
  \BibitemOpen
  \bibfield  {author} {\bibinfo {author} {\bibfnamefont {J.}~\bibnamefont {Wu}}, \bibinfo {author} {\bibfnamefont {M.}~\bibnamefont {Clementi}}, \bibinfo {author} {\bibfnamefont {E.}~\bibnamefont {Nitiss}}, \bibinfo {author} {\bibfnamefont {J.}~\bibnamefont {Hu}}, \bibinfo {author} {\bibfnamefont {C.}~\bibnamefont {Lafforgue}},\ and\ \bibinfo {author} {\bibfnamefont {C.-S.}\ \bibnamefont {Br{\`e}s}},\ }\bibfield  {title} {\bibinfo {title} {Bright and dark talbot pulse trains on a chip},\ }\href {https://doi.org/10.1038/s42005-023-01375-x} {\bibfield  {journal} {\bibinfo  {journal} {Commun. Phys.}\ }\textbf {\bibinfo {volume} {6}},\ \bibinfo {pages} {249} (\bibinfo {year} {2023})}\BibitemShut {NoStop}%
\bibitem [{\citenamefont {Br\"uggenj\"urgen}\ \emph {et~al.}(2026)\citenamefont {Br\"uggenj\"urgen}, \citenamefont {Fischer},\ and\ \citenamefont {Weitenberg}}]{bruggenjurgen_fischer_26}%
  \BibitemOpen
  \bibfield  {author} {\bibinfo {author} {\bibfnamefont {J.~C.}\ \bibnamefont {Br\"uggenj\"urgen}}, \bibinfo {author} {\bibfnamefont {M.~S.}\ \bibnamefont {Fischer}},\ and\ \bibinfo {author} {\bibfnamefont {C.}~\bibnamefont {Weitenberg}},\ }\bibfield  {title} {\bibinfo {title} {A phase microscope for quantum gases},\ }\href {https://doi.org/10.1126/science.adt1712} {\bibfield  {journal} {\bibinfo  {journal} {Science}\ }\textbf {\bibinfo {volume} {393}},\ \bibinfo {pages} {167} (\bibinfo {year} {2026})}\BibitemShut {NoStop}%
\bibitem [{\citenamefont {Rigol}(2005)}]{rigol_05}%
  \BibitemOpen
  \bibfield  {author} {\bibinfo {author} {\bibfnamefont {M.}~\bibnamefont {Rigol}},\ }\bibfield  {title} {\bibinfo {title} {Finite-temperature properties of hard-core bosons confined on one-dimensional optical lattices},\ }\href {https://doi.org/10.1103/PhysRevA.72.063607} {\bibfield  {journal} {\bibinfo  {journal} {Phys. Rev. A}\ }\textbf {\bibinfo {volume} {72}},\ \bibinfo {pages} {063607} (\bibinfo {year} {2005})}\BibitemShut {NoStop}%
\bibitem [{\citenamefont {Jordan}\ and\ \citenamefont {Wigner}(1928)}]{jordan_wigner_28}%
  \BibitemOpen
  \bibfield  {author} {\bibinfo {author} {\bibfnamefont {P.}~\bibnamefont {Jordan}}\ and\ \bibinfo {author} {\bibfnamefont {E.}~\bibnamefont {Wigner}},\ }\bibfield  {title} {\bibinfo {title} {{\"{U}}ber das paulische {\"a}quivalenzverbot},\ }\href {https://doi.org/10.1007/BF01331938} {\bibfield  {journal} {\bibinfo  {journal} {Z. Phys.}\ }\textbf {\bibinfo {volume} {47}},\ \bibinfo {pages} {631} (\bibinfo {year} {1928})}\BibitemShut {NoStop}%
\bibitem [{\citenamefont {Rigol}\ and\ \citenamefont {Muramatsu}(2004{\natexlab{a}})}]{rigol_muramatsu_04a}%
  \BibitemOpen
  \bibfield  {author} {\bibinfo {author} {\bibfnamefont {M.}~\bibnamefont {Rigol}}\ and\ \bibinfo {author} {\bibfnamefont {A.}~\bibnamefont {Muramatsu}},\ }\bibfield  {title} {\bibinfo {title} {Universal properties of hard-core bosons confined on one-dimensional lattices},\ }\href {https://doi.org/10.1103/PhysRevA.70.031603} {\bibfield  {journal} {\bibinfo  {journal} {Phys. Rev. A}\ }\textbf {\bibinfo {volume} {70}},\ \bibinfo {pages} {031603(R)} (\bibinfo {year} {2004}{\natexlab{a}})}\BibitemShut {NoStop}%
\bibitem [{\citenamefont {Rigol}\ and\ \citenamefont {Muramatsu}(2005{\natexlab{a}})}]{rigol_muramatsu_05a}%
  \BibitemOpen
  \bibfield  {author} {\bibinfo {author} {\bibfnamefont {M.}~\bibnamefont {Rigol}}\ and\ \bibinfo {author} {\bibfnamefont {A.}~\bibnamefont {Muramatsu}},\ }\bibfield  {title} {\bibinfo {title} {Ground-state properties of hard-core bosons confined on one-dimensional optical lattices},\ }\href {https://doi.org/10.1103/PhysRevA.72.013604} {\bibfield  {journal} {\bibinfo  {journal} {Phys. Rev. A}\ }\textbf {\bibinfo {volume} {72}},\ \bibinfo {pages} {013604} (\bibinfo {year} {2005}{\natexlab{a}})}\BibitemShut {NoStop}%
\bibitem [{\citenamefont {Rigol}\ and\ \citenamefont {Muramatsu}(2004{\natexlab{b}})}]{rigol_muramatsu_04b}%
  \BibitemOpen
  \bibfield  {author} {\bibinfo {author} {\bibfnamefont {M.}~\bibnamefont {Rigol}}\ and\ \bibinfo {author} {\bibfnamefont {A.}~\bibnamefont {Muramatsu}},\ }\bibfield  {title} {\bibinfo {title} {Emergence of quasicondensates of hard-core bosons at finite momentum},\ }\href {https://doi.org/10.1103/PhysRevLett.93.230404} {\bibfield  {journal} {\bibinfo  {journal} {Phys. Rev. Lett.}\ }\textbf {\bibinfo {volume} {93}},\ \bibinfo {pages} {230404} (\bibinfo {year} {2004}{\natexlab{b}})}\BibitemShut {NoStop}%
\bibitem [{\citenamefont {Rigol}\ and\ \citenamefont {Muramatsu}(2005{\natexlab{b}})}]{rigol_muramatsu_05b}%
  \BibitemOpen
  \bibfield  {author} {\bibinfo {author} {\bibfnamefont {M.}~\bibnamefont {Rigol}}\ and\ \bibinfo {author} {\bibfnamefont {A.}~\bibnamefont {Muramatsu}},\ }\bibfield  {title} {\bibinfo {title} {Free expansion of impenetrable bosons on one-dimensional optical lattices},\ }\href {https://doi.org/10.1142/S0217984905008876} {\bibfield  {journal} {\bibinfo  {journal} {Mod. Phys. Lett. B}\ }\textbf {\bibinfo {volume} {19}},\ \bibinfo {pages} {861} (\bibinfo {year} {2005}{\natexlab{b}})}\BibitemShut {NoStop}%
\bibitem [{\citenamefont {Xu}\ and\ \citenamefont {Rigol}(2017)}]{xu_rigol_17}%
  \BibitemOpen
  \bibfield  {author} {\bibinfo {author} {\bibfnamefont {W.}~\bibnamefont {Xu}}\ and\ \bibinfo {author} {\bibfnamefont {M.}~\bibnamefont {Rigol}},\ }\bibfield  {title} {\bibinfo {title} {Expansion of one-dimensional lattice hard-core bosons at finite temperature},\ }\href {https://doi.org/10.1103/PhysRevA.95.033617} {\bibfield  {journal} {\bibinfo  {journal} {Phys. Rev. A}\ }\textbf {\bibinfo {volume} {95}},\ \bibinfo {pages} {033617} (\bibinfo {year} {2017})}\BibitemShut {NoStop}%
\bibitem [{\citenamefont {Lieb}\ \emph {et~al.}(2004)\citenamefont {Lieb}, \citenamefont {Schultz},\ and\ \citenamefont {Mattis}}]{lieb_schultz_04}%
  \BibitemOpen
  \bibfield  {author} {\bibinfo {author} {\bibfnamefont {E.}~\bibnamefont {Lieb}}, \bibinfo {author} {\bibfnamefont {T.}~\bibnamefont {Schultz}},\ and\ \bibinfo {author} {\bibfnamefont {D.}~\bibnamefont {Mattis}},\ }\bibfield  {title} {\bibinfo {title} {Two soluble models of an antiferromagnetic chain},\ }in\ \href {https://doi.org/10.1007/978-3-662-06390-3_35} {\emph {\bibinfo {booktitle} {Condensed Matter Physics and Exactly Soluble Models: Selecta of Elliott H. Lieb}}},\ \bibinfo {editor} {edited by\ \bibinfo {editor} {\bibfnamefont {B.}~\bibnamefont {Nachtergaele}}, \bibinfo {editor} {\bibfnamefont {J.~P.}\ \bibnamefont {Solovej}},\ and\ \bibinfo {editor} {\bibfnamefont {J.}~\bibnamefont {Yngvason}}}\ (\bibinfo  {publisher} {Springer Berlin Heidelberg},\ \bibinfo {address} {Berlin, Heidelberg},\ \bibinfo {year} {2004})\ pp.\ \bibinfo {pages} {543--601}\BibitemShut {NoStop}%
\bibitem [{\citenamefont {Barouch}\ and\ \citenamefont {McCoy}(1971)}]{barouch_mccoy_71}%
  \BibitemOpen
  \bibfield  {author} {\bibinfo {author} {\bibfnamefont {E.}~\bibnamefont {Barouch}}\ and\ \bibinfo {author} {\bibfnamefont {B.~M.}\ \bibnamefont {McCoy}},\ }\bibfield  {title} {\bibinfo {title} {Statistical mechanics of the {$XY$} model. {II. S}pin-correlation functions},\ }\href {https://doi.org/10.1103/PhysRevA.3.786} {\bibfield  {journal} {\bibinfo  {journal} {Phys. Rev. A}\ }\textbf {\bibinfo {volume} {3}},\ \bibinfo {pages} {786} (\bibinfo {year} {1971})}\BibitemShut {NoStop}%
\bibitem [{\citenamefont {Bloch}\ \emph {et~al.}(2008)\citenamefont {Bloch}, \citenamefont {Dalibard},\ and\ \citenamefont {Zwerger}}]{bloch_dalibard_08}%
  \BibitemOpen
  \bibfield  {author} {\bibinfo {author} {\bibfnamefont {I.}~\bibnamefont {Bloch}}, \bibinfo {author} {\bibfnamefont {J.}~\bibnamefont {Dalibard}},\ and\ \bibinfo {author} {\bibfnamefont {W.}~\bibnamefont {Zwerger}},\ }\bibfield  {title} {\bibinfo {title} {Many-body physics with ultracold gases},\ }\href {https://doi.org/10.1103/RevModPhys.80.885} {\bibfield  {journal} {\bibinfo  {journal} {Rev. Mod. Phys.}\ }\textbf {\bibinfo {volume} {80}},\ \bibinfo {pages} {885} (\bibinfo {year} {2008})}\BibitemShut {NoStop}%
\bibitem [{\citenamefont {Surace}\ and\ \citenamefont {Tagliacozzo}(2022)}]{surace_tagiliacozzo_22}%
  \BibitemOpen
  \bibfield  {author} {\bibinfo {author} {\bibfnamefont {J.}~\bibnamefont {Surace}}\ and\ \bibinfo {author} {\bibfnamefont {L.}~\bibnamefont {Tagliacozzo}},\ }\bibfield  {title} {\bibinfo {title} {Fermionic {G}aussian states: an introduction to numerical approaches},\ }\href {https://doi.org/10.21468/SciPostPhysLectNotes.54} {\bibfield  {journal} {\bibinfo  {journal} {SciPost Phys. Lect. Notes}\ ,\ \bibinfo {pages} {54}} (\bibinfo {year} {2022})}\BibitemShut {NoStop}%
\bibitem [{\citenamefont {Grenander}\ and\ \citenamefont {Szeg{\H{o}}}(1958)}]{grenander_szego_58}%
  \BibitemOpen
  \bibfield  {author} {\bibinfo {author} {\bibfnamefont {U.}~\bibnamefont {Grenander}}\ and\ \bibinfo {author} {\bibfnamefont {G.}~\bibnamefont {Szeg{\H{o}}}},\ }\href@noop {} {\emph {\bibinfo {title} {Toeplitz Forms and Their Applications}}},\ \bibinfo {series} {California Monographs in Mathematical Sciences}, Vol.~\bibinfo {volume} {2}\ (\bibinfo  {publisher} {University of California Press},\ \bibinfo {address} {Berkeley},\ \bibinfo {year} {1958})\BibitemShut {NoStop}%
\bibitem [{\citenamefont {McCoy}\ and\ \citenamefont {Wu}(1968)}]{mccoy_wu_68}%
  \BibitemOpen
  \bibfield  {author} {\bibinfo {author} {\bibfnamefont {B.~M.}\ \bibnamefont {McCoy}}\ and\ \bibinfo {author} {\bibfnamefont {T.~T.}\ \bibnamefont {Wu}},\ }\bibfield  {title} {\bibinfo {title} {Theory of {T}oeplitz determinants and the spin correlations of the two-dimensional {I}sing model. {V}},\ }\href {https://doi.org/10.1103/PhysRev.174.546} {\bibfield  {journal} {\bibinfo  {journal} {Phys. Rev.}\ }\textbf {\bibinfo {volume} {174}},\ \bibinfo {pages} {546} (\bibinfo {year} {1968})}\BibitemShut {NoStop}%
\bibitem [{\citenamefont {B{\"o}ttcher}\ and\ \citenamefont {Silbermann}(1999)}]{bottcher_silbermann_99}%
  \BibitemOpen
  \bibfield  {author} {\bibinfo {author} {\bibfnamefont {A.}~\bibnamefont {B{\"o}ttcher}}\ and\ \bibinfo {author} {\bibfnamefont {B.}~\bibnamefont {Silbermann}},\ }\href {https://doi.org/10.1007/978-1-4612-1426-7} {\emph {\bibinfo {title} {Introduction to Large Truncated Toeplitz Matrices}}},\ Universitext\ (\bibinfo  {publisher} {Springer},\ \bibinfo {address} {New York},\ \bibinfo {year} {1999})\BibitemShut {NoStop}%
\end{thebibliography}%

\end{document}